# A crystallographic model of the {557} habit planes in low-carbon martensitic steels

Cyril Cayron*, Annick Baur, Roland Logé

Laboratory of ThermoMechanical Metallurgy (LMTM), PX Group Chair, Ecole Polytechnique Fédérale de Lausanne (EPFL), Rue de la Maladière 71b, 2000 Neuchâtel, Switzerland

* Corresponding author. Email: cyril.cayron@epfl.ch

## Abstract

Low-alloy steels are constituted of twenty-four variants of lath martensite that exhibit gradients of orientations from Kurdjumov-Sachs (KS) to Nishiyama-Wassermann (NW). They are structured into four packets on each of the common close-packed plane $\{111\}_\gamma$ // $\{110\}_\alpha$; and each packet is composed of three blocks constituted by pairs of low-misoriented variants. The habit planes reported in literature for this type of martensite are $\{557\}_\gamma$, but it is not clear whether they correspond to the laths or to the blocks. In this paper, we present crystallographic calculations proving that the average of the two KS distortions associated with the variants in a block is exactly a NW distortion. A new method of averaging distortion matrices was introduced for this purpose. It is also shown that the $\{575\}_\gamma$ planes are let untilted by this NW distortion, and are thus good theoretical candidates for the observed habit planes. The predicted $\{575\}_\gamma$ planes, however, do not contain any of the common close-packed directions of the two variants in the block, which is in apparent contradiction with the current view. In order to clarify this point, some Electron BackScatter Diffraction (EBSD) maps were acquired on different low-carbon steels; the prior austenitic grains were automatically reconstructed and the traces of the habit planes predicted by the different models were analyzed and compared to the morphologies. This experimental work shows that $\{575\}_\gamma$ planes are the habit plane of the blocks, and that the habit plane of one block often dominates the others, which impedes to discriminate the different models. The advantages of our model are its simplicity, the absence of fitting parameters, and the symmetric role played by the variants the blocks.

## 1. Introduction

Low-carbon steels and iron nickel alloys containing less than 28 wt-% Ni exhibit after quenching microstructures constituted of body centered cubic (bcc, $\alpha$) lath martensite. The mean orientation relationship (OR) with the face centered cubic austenite (fcc, $\gamma$) is close to Nishiyama-Wasserman (NW) [1][2], or in-between NW and Kurdjumov-Sachs (KS) [3], i.e. close to the Greninger-Troiano (GT) OR [4]. It is interesting to mention here the "hidden story" reported by Shimizu in his paper on Nishiyama's life: *"As known, angular difference of 5.27° exists between the Nishiyama's and the Kurdjumov-Sachs's orientation relationships. When he [Nishiyama] decided to publish the new orientation relationship and told the decision to Professor Honda, the boss advised him to reexamine the orientation relationship, pointing out a possibility that the angular difference of 5.27° might be within experimental error and the two orientation relationships might be essentially the same. However, Professor Nishiyama maintained the correctness of his new orientation relationship, emphasizing that the experimental error was much less than 5.27°..."* [5].

The martensitic laths are often small, with a thickness of few microns, and highly intricated, which makes their study by optical microscopy difficult. However, Marder and Krauss [6] elegantly took advantage that some long straight lines visible in the metallographies are twin boundaries in the prior austenitic grains to statistically determine the martensite habit planes; and they found that they are close to $\{557\}_\gamma$. Then, transmission electron microscopy (TEM) studies allowed observing the martensite at higher scales [7]-[11]. Some laths exhibit an exact KS OR but only locally, and in general the OR deviates from KS toward NW [8][10]. Later, the coupling of TEM with Electron Back Scatter Diffraction (EBSD) allowed the get a better idea of the structure of martensite at different scales. It is now established that the laths are grouped into four packets on the $\{111\}_\gamma$ // $\{110\}_\alpha$ close-packed planes. These packets are sometimes called morphological packets [12], or "close-packed plane" (CPP) packets [13]. Although the KS OR is only local, for sake of simplicity, it was admitted [10] that the twenty four variants can be indexed as if they were in KS OR. Morito *et al.* [11] showed that each packet contains six KS variants grouped by pairs of low-misoriented variants, called "blocks". The two KS variants in a block are linked by a rotation of 10.5° around the normal to the $(111)_\gamma$ // $(110)_\alpha$ plane, noted operator $O_5$ in Ref. [13]. The laths with close orientations are grouped in the blocks and form "sub-blocks". A schematic representation of the packets, blocks and sub-blocks is given by Morito *et al.* [11]. According to these authors, the laths have for habit plane the $\{557\}_\gamma$ planes that contain the dense direction $[\bar{1}10]_\gamma$ // $[\bar{1}11]_\alpha$, i.e. it is the plane $(557)_\gamma$ // $(156)_\alpha$. However, it should be kept in mind that the TEM works give only approximate results. Indeed, uncertainties come from the absence of retained austenite in low alloy steels, the existence of gradients of orientations inside the laths, and the fact that TEM lamella are often distorted due to the relaxation of the internal stresses, which means that the surface often deviates from the horizontality. In addition a logical problem appears. Let us consider two KS variants constituting a block, $\alpha_1$ and $\alpha_6$, as shown in Fig. 1; they share the common $(111)_\gamma$ // $(110)_\alpha$ plane but do not share a close-packed direction: $\alpha_1$ is such that $[1\bar{1}0]_\gamma$//$[\bar{1}11]_{\alpha 1}$ and $\alpha_6$ is such that $[0\bar{1}1]_\gamma$//$[\bar{1}11]_{\alpha 6}$. It would mean that the habit planes is $(557)_\gamma$ for the former and $(755)_\gamma$ for the latter. Therefore, one can wonder: what is the habit plane of the block? Morito's schematic representations of the intricate sub-block/block/packet microstructure remain elusive on this question. Krauss [14] solves the problem by assuming that the habit plane of one of the two variants in the block dominates, i.e. the habit plane of the block is either $(557)_\gamma$ or $(755)_\gamma$; however, in the TEM images in Ref. [7][11] show that the variants are in equal proportions, and in the TEM image of Fig.2 of Ref. [7] none of the two variants appear parallel to the block interface. Besides, if one of the two variants were really dominant in its block, it should be highly elongated along its dense direction because this direction is contained in the habit plane of the block, whereas the second variant should be very short; and this was not observed. All the TEM images show variants in equal proportion and with short lengths (few microns) in comparison with the block size (few hundreds of microns). Therefore, some questions related to lath martensite remain unsolved for the moment. Where does the spreading between the KS and the NW orientations come from? Are the {557} planes the habit planes of the nanometric laths, of the micronic blocks, or of both?

The phenomenological theory of martensitic crystallography (PTMC) failed to explain the {557} habit plane, at least in its early versions of 1953-54 [15][16], and it was nearly forty years later, in 1992, that Kelly [17] used the double-shear version of the PTMC developed by Ross and Croker [18] in order to make the theory more compatible with the experimental results. He used "floating" lattice invariant shear (LIS) systems with, for first LIS, shears of type $(hhl)[1\bar{1}0]_\gamma$ where the $(hhl)_\gamma$ planes varied from $(00\bar{1})_\gamma$ to $(110)_\gamma$, and for second LIS, shears on planes between $(311)_\gamma$ and $(3\bar{1}\bar{1})_\gamma$. He noticed that lath martensite with a habit plane close to $\{557\}_\gamma$ can be formed with at least 35 different combinations of LIS. His approach was used by Lambert-Perlade *et al.* [1] and Morito *et al.* [11] to discuss their EBSD results. They respectively chose $(\bar{2}31)[11\bar{1}]_\alpha$ & $(211)[1\bar{1}\bar{1}]_\alpha$ and $(112)[\bar{1}\bar{1}1]_\alpha$ & $(110)[\bar{1}11]_\alpha$, as double LIS systems. More recently, Iwashita *et al.* [19] used a PTMC version developed by Khachaturyan [20] with $(\bar{1}01)[101]\alpha$ & $(101)[\bar{1}01]_\alpha$ double-slip, and they adjusted the magnitude of shear on both planes to obtain results close to the experiments, i.e. an habit plane close to {557} and an orientation relationship close to KS. Later, Qi, Khachaturyan and Morris [21] followed a similar approach and explored all the combinations related simple and double slips with the variants chosen among the twenty-four KS variants and slips chosen among seventy-two systems. They tested more than 23'000 possibilities and concluded that the choice that gives a solution close to the experimental results is the composite case of two KS variants sharing the same Bain axis, i.e. the variants forming the blocks. To our point of view, the trial-and-error method and the high degree of freedom given to the LIS significantly reduce the "predictive" character of these PTMC approaches. In addition, to our knowledge, there is no clear experimental confirmation that the LIS proposed by the different authors fit with experimental observations, as those reported in TEM by Sandvik and Wayman [8].

For the last years, we have developed an alternative model to PTMC [22][23][24]. The analytical form of the atomic displacements and lattice distortion was determined only by assuming the orientation relationship and by considering the iron atoms as hard-spheres of constant size [24]. The concept of "natural" distortion imagined as the distortion freed from any stress was introduced. Great expectations were specially raised on the KS distortion because the $\{225\}_\gamma$ habit plane simply emerges from the angular distortion matrix as an untilted plane [24][25]. We could then prove that this plane is rendered fully invariant by combining two twin-related KS variants, i.e. misoriented by a rotation of 60° around the common close-packed direction $[\bar{1}10]_\gamma //[\bar{1}11]_\alpha$ [26]. The use of angular distortive matrices was also generalized to other martensitic transformations and some mechanical twinning in close-packed alloys [28][29]. The aim of the present paper is to show that the average distortion of the two KS variants in a block, i.e. the mean distortion of a block, is a NW distortion matrix, and that the $\{557\}_\gamma$ habit planes naturally emerge from this matrix. This predicted habit plane is at 9.4° from the common close-packed plane, but it does not contain the close-packed direction of any of the two KS variants. In the example of the block made of the variants $\alpha_1$ and $\alpha_6$, the predicted habit plane of the block is neither the $(557)_\gamma$ plane or the $(755)_\gamma$ plane; it is the $(575)_\gamma$ plane. It means that our model does not fit with the current point of view based on Morito's TEM study and the

Krauss' hypothesis. EBSD maps will be acquired on three low-carbon steels in the aim to find experimental evidences to discriminate the models. The EBSD maps will be analyzed with the software ARPGE [27] in order to reconstruct the austenite parent grains and plot the traces of the habit planes predicted by these models.

## 2. Experimental methods and crystallographic calculations

### 2.1. Experimental methods

The low-carbon steels chosen for this work are a low alloyed 1018 steel, a 8Cr-1W (Eurofer) steel and 9Cr-1Mo (EM10) steels. Their compositions are given in Table 1; their carbon content is lower than 0.2%. The steels were cut, sealed in a quartz tube under vacuum, heat treated at 1000°C for one hour, and water quenched. The samples were polished down to 1 micron and electropolished at 20V with the electrolyte A2 of Struers. The EBSD maps were acquired with a XLF30 (FEI) scanning electron microscope equipped with a Nordlys 2 camera and Aztec (Oxford Instruments) software. The acceleration voltage was 20 kV, the chosen magnifications were quite low (x500, x650 and x1200), and the step size between 0.1 and 0.3 microns. In the three steels, the only phase that could be identified is the bcc phase; there was no retained austenite. A special module has been developed in ARPGE to draw in the EBSD maps the trace of habit planes expected from the theoretical models. It can plot the traces of the equivalent planes $\{hkl\}_\alpha$ for each martensitic bcc grain, and the traces of the planes $\{hkl\}_\gamma$ for each parent grain; it can also plot in each bcc grains the traces of the planes $(hkl)_\alpha$ that are parallel to a plane of type $\{h'k'l'\}_\gamma$ with a certain tolerance angle fixed to 3°; for example all the traces of the daughter planes $\{110\}_\alpha$ such that $\{110\}_\alpha // \{111\}_\gamma$.

### 2.2. Reminder on the calculations of orientation and distortion variants

We note the reference basis of the γ phase $\mathbf{B}_0^\gamma = (\mathbf{a}_0^\gamma, \mathbf{b}_0^\gamma, \mathbf{c}_0^\gamma)$ with $\mathbf{a}_0^\gamma = [100]_\gamma$, $\mathbf{b}_0^\gamma = [010]_\gamma$, $\mathbf{c}_0^\gamma = [001]_\gamma$. The distortion matrix is formed by the coordinates of the transformed vectors $(\mathbf{a}_0^\gamma)'$, $(\mathbf{b}_0^\gamma)'$, $(\mathbf{c}_0^\gamma)'$ expressed in the reference basis $\mathbf{B}_0^\gamma$, written by

$$\mathbf{D}^{\gamma \to \alpha} = [\mathbf{B}_0^\gamma \to \mathbf{B}_0^{\gamma'}] \qquad (1)$$

This matrix can be calculated by finding a primitive basis of the parent phase and by following how this basis is distorted during the transformation [24][28]. The distortion matrix gives the images of the directions. The images of the planes are given by

$$(\mathbf{D}^{\gamma \to \alpha})^* = (\mathbf{D}^{\gamma \to \alpha})^{-T} \qquad (2)$$

where "-T" means "inverse of the transpose".

We also call $\mathbf{T}^{\gamma \to \alpha}$ the matrix of coordinate transformation, which gives the coordinates of the vectors of the reference basis $\mathbf{B}_0^\alpha$ of the α crystal in the reference basis $\mathbf{B}_0^\gamma$ of the γ crystal:

$$\mathbf{T}^{\gamma \to \alpha} = [\mathbf{B}_0^\gamma \to \mathbf{B}_0^\alpha] \qquad (3)$$

$\mathbf{T}^{\gamma \to \alpha}$ depends only on the orientation relationship. It can be used to write the vectors transformed by the distortion $(\mathbf{a}_0^\gamma)'$, $(\mathbf{b}_0^\gamma)'$, $(\mathbf{c}_0^\gamma)'$ in the reference basis $\mathbf{B}_0^\alpha$ of the α daughter phase; and the set of images forms the "correspondence" matrix, noted $\mathbf{C}_0^{\gamma \to \alpha}$, given by

$$\mathbf{C}^{\alpha \to \gamma} = \mathbf{T}^{\alpha \to \gamma} \mathbf{D}^{\gamma \to \alpha} = (\mathbf{T}^{\gamma \to \alpha})^{-1} \mathbf{D}^{\gamma \to \alpha} \qquad (4)$$

As discussed in Ref. [28], a distinction should be made between the orientational variants and the distortional variants. The orientational variants depend only on the orientation relationship and not on the mechanism. They can be mathematically defined from the subgroup $\mathbb{H}^\gamma$ of the *crystallographic* symmetries that are common to both parent and daughter crystals. This subgroup is given by the orientation matrix $\mathbf{T}^{\gamma \to \alpha}$ and by the point groups of parent and daughter phases, $\mathbb{G}^\gamma$ and $\mathbb{G}^\alpha$:

$$\mathbb{H}^\gamma = \mathbb{G}^\gamma \cap \mathbf{T}^{\gamma \to \alpha} \mathbb{G}^\alpha (\mathbf{T}^{\gamma \to \alpha})^{-1} \qquad (5)$$

The orientational variants are the cosets $\alpha_i = g_i^\gamma \mathbb{H}^\gamma$ and their orientation matrices are $\mathbf{T}^{\gamma \to \alpha_i} = g_i^\gamma \mathbf{T}^{\gamma \to \alpha}$ with $g_i^\gamma \in \alpha_i$. The number of orientational variants is $N^\alpha = \frac{|\mathbb{G}^\gamma|}{|\mathbb{H}^\gamma|}$. Details are given in Ref.[30].

The distortional variants depend on the mechanism, and more specifically on the distortion matrix. They can be defined similarly as for the orientational variants [28], but by replacing the orientation matrix $\mathbf{T}^{\gamma \to \alpha_i}$ by the distortion matrix $\mathbf{D}^{\gamma \to \alpha}$. The subgroup $\mathbb{K}^\gamma$ of *shape* symmetries that are common to both parent and daughter crystals is

$$\mathbb{K}^\gamma = \mathbb{G}^\gamma \cap \mathbf{D}^{\gamma \to \alpha} \mathbb{G}^\gamma (\mathbf{D}^{\gamma \to \alpha})^{-1} \qquad (6)$$

The distortional variants are the cosets $d_i = g_i^\gamma \mathbb{K}^\gamma$ and their distortion matrices are $\mathbf{D}^{\gamma \to \alpha_i} = g_i^\gamma \mathbf{D}^{\gamma \to \alpha} (g_i^\gamma)^{-1}$ with $g_i^\gamma \in d_i$. The number of distortional variants is $M^\alpha = \frac{|\mathbb{G}^\gamma|}{|\mathbb{K}^\gamma|}$.

The distortion matrices associated with the KS OR and with the NW OR will be noted $\mathbf{D}^{KS}$ and $\mathbf{D}^{NW}$, and simply called here "KS distortion matrix" and "NW distortion matrix", respectively. However, it is important to keep in mind that these matrices are not those proposed by Kurdjumov and Sachs [31] and by Nishiyama [32], respectively. Indeed, in their seminal papers often cited for the experimental determination of the orientation relationship, the researchers also proposed alternatives to the Bain model that suppose that the lattice deformation results from a simple shear combined with an expansion and a contraction along some crystallographic directions.

Please also note that the distortion matrix $\mathbf{D}^{\gamma \to \alpha}$ was noted $\mathbf{D}_0^{\gamma \to \alpha}$ in ref. [24][28], the index "0" meaning the reference basis $\mathbf{B}_0^\gamma$. In the present work, the index $i$ is attributed to the index of the distortional variant, i.e. it is the matrix of the distortional variant $i$ obtained with the KS orientation relationship, and noted for sake of simplicity $\mathbf{D}^{\gamma \to \alpha_i} = \mathbf{D}_i^{KS}$. By default, the identity matrix is $g_1^\gamma$, and the distortion matrix obtained with $g_1^\gamma$ (or any matrix in $\mathbb{K}^\gamma$) is noted $\mathbf{D}_1^{KS}$.

## 3. Calculations of the KS and NW distortion matrices

The KS OR that we used in our previous works [24][28] was $(\bar{1}11)_\gamma \, // \, (\bar{1}10)_\alpha$ & $[110]_\gamma = [111]_\alpha$. This choice was made to point out the importance of the invariant line $[110]_\gamma = [111]_\alpha$. Another equivalent choice is often made: $(111)_\gamma \, // \, (110)_\alpha$ & $[\bar{1}10]_\gamma = [\bar{1}11]_\alpha$. In order to facilitate the comparison with literature, this OR will be chosen in the present paper. The fcc-bcc distortion matrices can be calculated from $\mathbf{D}^{KS}$ given [24][28] by using the symmetry matrix $g = \begin{bmatrix} 0 & -1 & 0 \\ 1 & 0 & 0 \\ 0 & 0 & 1 \end{bmatrix}$, and by applying the coordinate transformation $g^{-1}\mathbf{D}^{KS}g$. It becomes

$$\mathbf{D}_1^{KS} = \begin{pmatrix} 1 + \dfrac{\sqrt{6}}{18} & \dfrac{\sqrt{6}}{18} & \dfrac{1}{3} - \dfrac{\sqrt{6}}{6} \\ -\dfrac{1}{3} + \dfrac{\sqrt{6}}{18} & \dfrac{2}{3} + \dfrac{\sqrt{6}}{18} & \dfrac{1}{3} - \dfrac{\sqrt{6}}{6} \\ \dfrac{1}{3} - \dfrac{\sqrt{6}}{9} & \dfrac{1}{3} - \dfrac{\sqrt{6}}{9} & \dfrac{1}{3} + \dfrac{\sqrt{6}}{3} \end{pmatrix} \quad (7)$$

The calculations prove that the intersection groups $\mathbb{H}^\gamma$ and $\mathbb{K}^\gamma$ are equal; both are constituted of the identity and inversion elements. Therefore, the orientational and distortional variants are equivalent, and one can associate to each distortion matrix $\mathbf{D}_i^{KS}$ a unique orientational variant $\alpha_i^{KS}$. Among the twenty-four KS distortion matrices, two are of particular interest relatively to $\mathbf{D}_1^{KS}$; they are $\mathbf{D}_3^{KS}$ and $\mathbf{D}_6^{KS}$. The matrix $\mathbf{D}_3^{KS}$ gives the variant $\alpha_3^{KS}$ linked to $\alpha_1^{KS}$ by the operator $O_1$ [13] (they are twin-related), and their association makes fully invariant the $\{225\}_\gamma$ plane [26]. The matrix $\mathbf{D}_6^{KS}$ gives the variant $\alpha_6^{KS}$ that also shares the same $(111)_\gamma // (110)_\alpha$ plane as $\alpha_1^{KS}$ and $\alpha_3^{KS}$. As illustrated in Fig. 1b, the variants $\alpha_1^{KS}$ and $\alpha_6^{KS}$ are linked by a rotation of 10.5° around the $[111]_\gamma // [110]_\alpha$ axis, i.e. by the operator $O_5$ [13]. The distortion matrix $\mathbf{D}_6^{KS}$ associated with the variant $\alpha_6^{KS}$ is deduced from $\mathbf{D}_1^{KS}$ by using the symmetry matrix $g_6^\gamma = \begin{pmatrix} 0 & 0 & 1 \\ 0 & 1 & 0 \\ 1 & 0 & 0 \end{pmatrix}$.

$$\mathbf{D}_6^{KS} = g_6^\gamma \mathbf{D}_1^{KS}(g_6^\gamma)^{-1} = \begin{pmatrix} \dfrac{1}{3} + \dfrac{\sqrt{6}}{3} & \dfrac{1}{3} - \dfrac{\sqrt{6}}{9} & \dfrac{1}{3} - \dfrac{\sqrt{6}}{9} \\ \dfrac{1}{3} - \dfrac{\sqrt{6}}{6} & \dfrac{2}{3} + \dfrac{\sqrt{6}}{18} & -\dfrac{1}{3} + \dfrac{\sqrt{6}}{18} \\ \dfrac{1}{3} - \dfrac{\sqrt{6}}{6} & \dfrac{\sqrt{6}}{18} & 1 + \dfrac{\sqrt{6}}{18} \end{pmatrix} \quad (8)$$

It can be checked that $\mathbf{D}_1^{KS}$ and $\mathbf{D}_6^{KS}$ let invariant the directions $[1\bar{1}0]_\gamma$ and $[01\bar{1}]_\gamma$, respectively. It will be shown in the next section that the average of $\mathbf{D}_1^{KS}$ and $\mathbf{D}_6^{KS}$ is a distortion that lets unrotated the third close-packed direction of the $(111)_\gamma$ plane, i.e. the $[10\bar{1}]_\gamma$ direction, and that is exactly equal to the distortion related to the NW OR. For this reason, the block formed by the variants $\alpha_1^{KS}$ and $\alpha_6^{KS}$ variants is noted $\alpha_1^{NW}$.

For the moment, one needs to calculate the NW distortion matrix by another method. We use the fact that the NW variant $\alpha_1^{NW}$ with orientation $(111)_\gamma // (110)_\alpha$ and $[10\bar{1}]_\gamma // [001]_\alpha$ is located at 5.26° between the two variants $\alpha_1^{KS}$ and $\alpha_6^{KS}$, as illustrated in Fig.1b. The rotation matrix that re-orients $\alpha_1^{KS}$ to $\alpha_1^{NW}$ is the rotation $\mathbf{R} = \mathbf{R}_{(\alpha_1^{KS} \to \alpha_1^{NW})}$ of 5.26° around $[111]_\gamma$, and the one that that re-orients $\alpha_6^{KS}$ to $\alpha_1^{NW}$ is the rotation $\mathbf{R}^{-1} = \mathbf{R}_{(\alpha_6^{KS} \to \alpha_1^{NW})}$ of -5.26° around $[111]_\gamma$, with:

$$\mathbf{R} = \begin{pmatrix} \tfrac{1}{9}(3 + 3\sqrt{2} + \sqrt{3}) & \tfrac{1}{9}(3 - 2\sqrt{3}) & \tfrac{1}{9}(3 - 3\sqrt{2} + \sqrt{3}) \\ \tfrac{1}{9}(3 - 3\sqrt{2} + \sqrt{3}) & \tfrac{1}{9}(3 + 3\sqrt{2} + \sqrt{3}) & \tfrac{1}{9}(3 - 2\sqrt{3}) \\ \tfrac{1}{9}(3 - 2\sqrt{3}) & \tfrac{1}{9}(3 - 3\sqrt{2} + \sqrt{3}) & \tfrac{1}{9}(3 + 3\sqrt{2} + \sqrt{3}) \end{pmatrix} \quad (9)$$

The distortion matrix associated with the variant $\alpha_1^{NW}$ is thus

$$\mathbf{D}_1^{NW} = \mathbf{R}\mathbf{D}_1^{KS} = \mathbf{R}^{-1}\mathbf{D}_6^{KS} = \begin{pmatrix} \frac{1}{6}(2+\sqrt{2}+2\sqrt{3}) & \frac{1}{6}(2-\sqrt{2}) & \frac{1}{6}(2+\sqrt{2}-2\sqrt{3}) \\ \frac{1}{3}(1-\sqrt{2}) & \frac{1}{3}(1+\sqrt{2}) & \frac{1}{3}(1-\sqrt{2}) \\ \frac{1}{6}(2+\sqrt{2}-2\sqrt{3}) & \frac{1}{6}(2-\sqrt{2}) & \frac{1}{6}(2+\sqrt{2}+2\sqrt{3}) \end{pmatrix} \quad (10)$$

It can be checked that this matrix lets unrotated the $[10\bar{1}]_\gamma$ direction. This direction is just elongated by a ratio of $2\sqrt{3}/3 \approx 1.155$. Further calculations (not detailed here) prove that here again there are as many NW orientational variants as distortional ones. Therefore, a one-to-one correspondence between the orientational variants and the distortional variants is again appropriate. The polar decomposition of $\mathbf{D}_1^{NW}$ will be used in the next section. It is written

$$\mathbf{D}_1^{NW} = \mathbf{R}_1^{NW}\mathbf{B}_y \quad (11)$$

$$= \begin{pmatrix} \frac{1}{12}(6+2\sqrt{3}+\sqrt{6}) & \frac{-1+\sqrt{2}}{2\sqrt{3}} & \frac{1}{12}(-6+2\sqrt{3}+\sqrt{6}) \\ \frac{1}{6}(\sqrt{3}-\sqrt{6}) & \frac{1}{\sqrt{3}}+\frac{1}{\sqrt{6}} & \frac{1}{6}(\sqrt{3}-\sqrt{6}) \\ \frac{1}{12}(-6+2\sqrt{3}+\sqrt{6}) & \frac{-1+\sqrt{2}}{2\sqrt{3}} & \frac{1}{12}(6+2\sqrt{3}+\sqrt{6}) \end{pmatrix} \begin{pmatrix} \frac{2}{\sqrt{3}} & 0 & 0 \\ 0 & \sqrt{\frac{2}{3}} & 0 \\ 0 & 0 & \frac{2}{\sqrt{3}} \end{pmatrix}$$

The rotation $\mathbf{R}_1^{NW}$ is of angle $\mathrm{ArcCos}(\frac{1}{3}+\frac{1}{\sqrt{6}})$ and of axis $[\bar{1}01]_\gamma$, and $\mathbf{B}_y$ is the classical Bain distortion matrix with the contraction axis oriented along the $[010]_\gamma$ direction.

## 4. The NW distortion matrix as the average of two KS distortion matrices

By considering how the vectors are transformed by $\mathbf{D}_1^{KS}$ and $\mathbf{D}_6^{KS}$, one gets the geometrical feeling that the averaging of these two KS distortion matrices should lead to the $\mathbf{D}_1^{NW}$ matrix. How can it be rigorously proved? The more general question is: how to average two distortion matrices $\mathbf{D}_1$ and $\mathbf{D}_2$? The classical method used in the PTMC [16][20][33] consists in linearly adding the distortion matrices with a weight factor $0 \leq \lambda \leq 1$ such that the resulting distortion matrix is assumed to be

$$\mathbf{L} = \lambda\mathbf{D}_1 + (1-\lambda)\mathbf{D}_2 \quad (12)$$

If one applies this formula to combine with equal weight the two distortion matrices $\mathbf{D}_1^{KS}$ and $\mathbf{D}_6^{KS}$, it comes $\mathbf{L} = \frac{1}{2}\mathbf{D}_1^{KS} + \frac{1}{2}\mathbf{D}_6^{KS}$. By considering the exact values given in equations (7) and (8), and by equation (10), it is obvious that $\mathbf{L}$ is not exactly equal to $\mathbf{D}_1^{NW}$. However, the values are very close; indeed, the numerical expressions of the matrices chopped at $10^{-4}$ are:

$$\mathbf{L} \approx \begin{pmatrix} 1.14296 & 0.0986253 & -0.0068736 \\ -0.136083 & 0.802749 & -0.136083 \\ -0.0068736 & 0.0986253 & 1.14296 \end{pmatrix}, \mathbf{D}_1^{NW} \approx \begin{pmatrix} 1.14639 & 0.0976311 & -0.0083147 \\ -0.138071 & 0.8047378 & -0.138071 \\ -0.0083147 & 0.0976311 & 1.14639 \end{pmatrix}$$

Why linear formula (12) is only approximate (at $5.10^{-3}$) and does not work exactly? The fundamental reason is that this formula does not respect the invariance of the determinant. Let us explain. The determinant of a distortion matrix gives the ratio of the atomic volume of the daughter phase divided by the atomic volume of the parent phase, i.e. the density of the parent phase divided by the density of the daughter phase. For fcc-bcc transformations with hard-sphere assumptions, this ratio is $(4/3)\sqrt{2/3} \approx 1.08866$. Combining two distortion matrices should not modify this ratio because it is a volume property. However, the determinant is *not* a linear function of matrices, which means that in general $\det(\mathbf{L}) \neq \lambda.\det(\mathbf{D}_1) + (1-\lambda).\det(\mathbf{D}_2)$. Thus, although often used, formula (12) is most of the time incorrect because it does not conserve the volume.

Does an averaging formula leading exactly to the NW distortion matrix exist? The answer is positive. The idea lies in the fact that both $\alpha_1^{KS}$ and $\alpha_6^{KS}$ are two variants slightly misoriented; they belong to the same Bain packet; their <100> directions form the same <100> Bain circle in the pole figure; they have the same Bain matrix in their polar decomposition. Both distortion matrices $\mathbf{D}_1^{KS}$ and $\mathbf{D}_6^{KS}$ can indeed be decomposed as a product of a rotation and the Bain matrix $\mathbf{B}_y$ :

$$\mathbf{D}_1^{KS} = \mathbf{R}_1^{KS}\mathbf{B}_y = \begin{pmatrix} \frac{1}{12}(\sqrt{2}+6\sqrt{3}) & \frac{1}{6} & -\frac{-2+\sqrt{6}}{4\sqrt{3}} \\ \frac{-6+\sqrt{6}}{12\sqrt{3}} & \frac{1}{6}+\sqrt{\frac{2}{3}} & -\frac{-2+\sqrt{6}}{4\sqrt{3}} \\ \frac{1}{6}\sqrt{5-2\sqrt{6}} & -\frac{1}{3}+\frac{1}{\sqrt{6}} & \sqrt{\frac{7}{12}+\frac{1}{\sqrt{6}}} \end{pmatrix} \begin{pmatrix} \frac{2}{\sqrt{3}} & 0 & 0 \\ 0 & \sqrt{\frac{2}{3}} & 0 \\ 0 & 0 & \frac{2}{\sqrt{3}} \end{pmatrix} \quad (13)$$

$$\mathbf{D}_6^{KS} = \mathbf{R}_6^{KS}\mathbf{B}_y = \begin{pmatrix} \sqrt{\frac{7}{12}+\frac{1}{\sqrt{6}}} & -\frac{1}{3}+\frac{1}{\sqrt{6}} & \frac{1}{6}\sqrt{5-2\sqrt{6}} \\ -\frac{-2+\sqrt{6}}{4\sqrt{3}} & \frac{1}{6}+\sqrt{\frac{2}{3}} & \frac{-6+\sqrt{6}}{12\sqrt{3}} \\ -\frac{-2+\sqrt{6}}{4\sqrt{3}} & \frac{1}{6} & \frac{1}{12}(\sqrt{2}+6\sqrt{3}) \end{pmatrix} \begin{pmatrix} \frac{2}{\sqrt{3}} & 0 & 0 \\ 0 & \sqrt{\frac{2}{3}} & 0 \\ 0 & 0 & \frac{2}{\sqrt{3}} \end{pmatrix} \quad (14)$$

The rotations $\mathbf{R}_1^{KS}$ and $\mathbf{R}_6^{KS}$ are of angle $\text{ArcCos}(-\frac{5}{12}+\frac{7}{12\sqrt{2}}+\frac{1}{\sqrt{3}}+\frac{1}{\sqrt{6}})$, and their rotation axes are $[1-\sqrt{2}, 7-5\sqrt{2}-4\sqrt{3}+3\sqrt{6}, 1]$ and $[1, 7-5\sqrt{2}-4\sqrt{3}+3\sqrt{6}, 1-\sqrt{2}]$, respectively. The distortion matrices can be averaged while maintaining the determinant invariant; this is done by averaging the rotation matrices $\mathbf{R}_1^{KS}$ and $\mathbf{R}_6^{KS}$, and then by multiplying the mean rotation by the Bain matrix $\mathbf{B}_y$. The question is now: how to average two rotation matrices? This question is not simple and has been the subject of a broad literature well beyond material science. There are many formulae that can be applied depending on the chosen metrics (geodesic, quaternion, chordal etc.) [34][35]. The average of two rotations should give a rotation, which implies the invariance of determinant. Besides, we search a formula in which the two rotation matrices can be interchanged in order to remove the non-commutativity problem that comes if one adopts a naïve direct geometric averaging (taking the square root of a matrix product depends on the order of the matrices in the product). Such a remarkable formula exists; it was proposed by Moahker in 2002 [36]: the geometric mean of two rotations $\mathbf{R}_1$ and $\mathbf{R}_2$ is

$$\langle \mathbf{R}_1, \mathbf{R}_2 \rangle = \mathbf{R}_1(\mathbf{R}_1^T\mathbf{R}_2)^{1/2} = \mathbf{R}_2(\mathbf{R}_2^T\mathbf{R}_1)^{1/2} = \langle \mathbf{R}_2, \mathbf{R}_1 \rangle \quad (15)$$

This formula is of prime importance, although not yet well known in crystallography and metallurgy. If two distortion matrices $\mathbf{D}_1$ and $\mathbf{D}_2$ are based on the same symmetric matrix $\mathbf{B}$, i.e. $\mathbf{D}_1 = \mathbf{R}_1\mathbf{B}$ and $\mathbf{D}_2 = \mathbf{R}_2\mathbf{B}$, their mean is thus given by

$$\langle \mathbf{D}_1, \mathbf{D}_2 \rangle = \langle \mathbf{R}_1, \mathbf{R}_2 \rangle \mathbf{B} \tag{16}$$

Applied to the KS rotations $\mathbf{R}_1^{KS}$ and $\mathbf{R}_6^{KS}$ it gives

$$\langle \mathbf{R}_1^{KS}, \mathbf{R}_6^{KS} \rangle = \begin{pmatrix} \frac{1}{12}(6+2\sqrt{3}+\sqrt{6}) & -\frac{1}{6}(\sqrt{3}-\sqrt{6}) & \frac{1}{12}(-6+2\sqrt{3}+\sqrt{6}) \\ \frac{1}{6}(\sqrt{3}-\sqrt{6}) & \frac{1}{\sqrt{3}}+\frac{1}{\sqrt{6}} & \frac{1}{6}(\sqrt{3}-\sqrt{6}) \\ \frac{1}{12}(-6+2\sqrt{3}+\sqrt{6}) & -\frac{1}{6}(\sqrt{3}-\sqrt{6}) & \frac{1}{12}(6+2\sqrt{3}+\sqrt{6}) \end{pmatrix} \tag{17}$$

The matrix is exactly the rotation obtained by the polar decomposition of $\mathbf{D}_1^{NW}$ in equation (11). The matrix $\mathbf{B}$ is $\mathbf{B}_y$ of equations (11),(13) and (14). Thus, we get exactly the equation that was expected:

$$\langle \mathbf{D}_1^{KS}, \mathbf{D}_6^{KS} \rangle = \mathbf{D}_1^{NW} \tag{18}$$

The calculations prove that average of the distortion matrices related to KS variants in a block is equal to a NW distortion matrix. This theoretical result is in good agreement with the fact that locally the orientation relationship of a lath can be close KS, but that more globally; it tends to be closer to NW.

## 5. Habit planes of the blocks: predictions

The average distortion matrix related to the two KS variants in a block is a NW distortion matrix. The question now is: can the observed $\{557\}_\gamma$ habit planes be explained by this NW distortion matrix, for example by using with the "untilted-plane" criterion as we have done for the {225} habit planes in the high-carbon steels [24][25]? If it is the hypothesis is correct, the habit plane of a block should be untilted by the NW distortion, i.e. it should be equal to one of the eigenvectors of the NW distortion matrix expressed in the reciprocal space:

$$(\mathbf{D}_1^{NW})^* = (\mathbf{D}_1^{NW})^{-T} = \begin{pmatrix} \frac{1}{8}(2+\sqrt{2}+2\sqrt{3}) & \frac{1}{4}(2-\sqrt{2}) & \frac{1}{8}(2+\sqrt{2}-2\sqrt{3}) \\ \frac{1}{4}(1-\sqrt{2}) & \frac{1}{2}+\frac{1}{\sqrt{2}} & \frac{1}{4}(1-\sqrt{2}) \\ \frac{1}{8}(2+\sqrt{2}-2\sqrt{3}) & \frac{1}{4}(2-\sqrt{2}) & \frac{1}{8}(2+\sqrt{2}+2\sqrt{3}) \end{pmatrix} \tag{19}$$

This matrix has for eigenvalues $\frac{3}{2\sqrt{2}} \approx 1.0607$, 1 and $\frac{\sqrt{3}}{2} \approx 0.8660$, and for respective eigenvectors the planes $(1,\sqrt{2},1)_\gamma$, $(1,1,1)_\gamma$ and $(\bar{1},0,1)_\gamma$.

We have concluded in Ref. [24] that $(111)_\gamma$ cannot be a habit plane for martensite because it is the plane in which the fcc-bcc volume change is "localized". Indeed, in the ideal case of hard-sphere model, the fcc-bcc volume change only comes from the increase of angle between two close-packed <110>$_\gamma$ directions in the $(111)_\gamma$ plane transformed into $(110)_\alpha$ (Fig. 1c); as this angle increases from 60° to 70.5° while the lengths of the two directions remain constant, the surface increases by $\frac{4}{3}\sqrt{\frac{2}{3}} \approx 1.088$, which is exactly the fcc-bcc volume change. To our knowledge, the only case of $\{111\}_\gamma$ habit planes concerns the iron "octahedrite" meteorites (see **Supplementary Material 1**), but it is probable that the very low cooling rates of these objects (few degrees per thousand years) allowed diffusive accommodation mechanisms, such as dislocation climbs, that usually can't be activated during rapid cooling.

The two remaining candidates for habit planes of the blocks are the $(1\sqrt{2}\,1)_\gamma$ and $(\bar{1}01)_\gamma$ planes. Even if the reasons are not clear, it seems that very often the habit plane chosen by martensite is the irrational one. Here, it is the plane is $(1,\sqrt{2},1)_\gamma$. This plane is at 0.3° of the experimental $(575)_\gamma$ plane. Is it a casual coincidence? We don't think so. It should be noted that this plane does not contain either of the close packed directions $[\bar{1}10]_\gamma$ or $[0\bar{1}1]_\gamma$ that are let invariant by the distortion $\mathbf{D}_1^{KS}$ or $\mathbf{D}_6^{KS}$, respectively. This result was expected from the symmetries; indeed, there is no reason that the average of two distortions in equal proportion favors one of the two variants. However, this result is striking because it would mean that, contrarily to what is currently admitted, the $\{557\}_\gamma$ habit planes of the blocks do not correspond to any of the two KS lath variants. This would imply that the habit planes of the laths determined by TEM are different from the {557} habit planes of the blocks determined by optical microscopy. In order to forge our own opinion on the nature of the habit planes of the blocks, we decided to study them by EBSD in the three low-carbon steels (section 2.1).

## 6. Habit planes of the blocks: EBSD experiments

### 6.1. Parent grain reconstruction

The indexation rates of the EBSD maps acquired in the present work are between 65% and 75%, and all the maps have been dilated in order to remove the unindexed pixels and put in contact the neighboring grains. The software ARPGE was used to automatically reconstruct the prior parent grains and their orientations with the same parameters. An angular tolerance of 5° was chosen to identify the daughter bcc grains; the NW OR was used for the austenite reconstruction, with tolerance angles starting at 3° ("nucleation" step) and increasing up to 15°. Using the NW OR allows fast reconstructions (few minutes) because of this OR involves only 12 variants and it works well for low-carbon steels, but it gives poor results in high-carbon steels, and KS or GT ORs should be preferred in these alloys. Once reconstructed, the orientations of the parent grains are used to index the bcc grains as if they were KS variants. The indices of the variants were then used to identify the crystallographic CPP packets. The validity of the reconstruction was checked by plotting the pole figures of the daughter bcc grains for some prior austenitic grains. These pole figures should exhibit special features, such as <100>$_\alpha$ circles and small two-fold bow ties, <110>$_\alpha$ three-fold stars and four-fold crosses, and large <111>$_\alpha$ two-fold bow ties. Examples of such features are given in Ref. [22][23]. The quality of the determination of the orientations of the parent grains can also be checked. The <100>$_\gamma$ reconstructed directions should appear in the centers of the <100>$_\alpha$ circles, the <111>$_\gamma$ directions should be in the centers of the <110>$_\alpha$ three-fold stars, and the <110>$_\gamma$ directions sguld be in the middles of the <111>$_\alpha$ bow ties. A simulation that illustrates one of these correspondences is given Fig. 2. These continuous features show the continuities between the Pitsch, KS and NW ORs that come from internal gradients, and which, to our point of view, should be consider as a plastic mark of fcc-bcc distortion mechanism [22][23][24]. These gradients make the low-misorientation operators, i.e. the $O_3$ (10.5° around <111>$_\alpha$) and $O_5$ (10.5° around <110>$_\alpha$) that link the variant $\alpha_1^{KS}$ to the variant $\alpha_4^{KS}$ and to the variant $\alpha_6^{KS}$, respectively (Fig. 1), nearly impossible to distinguish in the experimental EBSD maps of the low-carbon steels. The average orientation between $\alpha_1^{KS}$ and $\alpha_6^{KS}$ is the NW OR, and the average orientation between $\alpha_1^{KS}$ and $\alpha_4^{KS}$ is the Pitsch OR (see Ref. [23] for the details).

### 6.2. The {557} habit planes

We have seen in introduction that Morito *et al.* [11] consider the $\{557\}_\gamma$ habit planes as a property of the individual laths observed by TEM. However, Marder and Krauss [6] measured these habit planes by optical microscopy, i.e. at magnifications hundred times lower than TEM; which means that the $\{557\}_\gamma$ habit planes were those of the blocks (and not those of the laths). In order to keep a coherency between the different scales, Krauss made the hypothesis that a group of laths with the same $\{557\}_\gamma$ habit plane dominates on the other groups and imposes its habit plane [14]. However, this hypothesis is not coherent with the fact that the variants appear in equal proportions in the TEM images reported in literature (see section 1). Our theoretical calculations show that the $\{557\}_\gamma$ habit plane of a block can be explained by a equi-proportion of the two KS variants, and that this habit plane should be different from that of any of the KS lath variants in the block. According to Morito et al. and Krauss' point of view, called in the rest of the paper "current" model, the $\{557\}_\gamma$ habit plane should be parallel to a $\{156\}_\alpha$ plane if one assumes a local KS OR with the parent austenite. In our model, the $\{557\}_\gamma$ habit planes should be parallel to a $\{750\}_\alpha$ plane by assuming a NW OR, or to $\{371, 265, 7\}_\alpha$ planes by assuming a KS OR. This is summarized by the conditions (a) and (b):

(a) Current model: Habit plane = $\{156\}_\alpha$ // $\{557\}_\gamma$
(b) Our model:     Habit plane = $\{750\}_\alpha$ // $\{557\}_\gamma$

Both hypotheses have been studied by comparing the traces of the predicted habit planes in the EBSD maps of low-carbon steels with a tolerance of 3° chosen for both conditions. All the maps were treated with the same parameters and in the same way. An example of analysis is given for the 1018 steel in Fig. 3. First, the initial EBSD map of the martensitic grains (Fig. 3a) is used to reconstruct the prior fcc austenitic grains with ARPGE (Fig. 3b), then the bcc martensitic grains are indexed as KS variants, and their internal misorientations ranging from Pitsch to NW are colored with a RGB code (Red = Pitsch, Green = KS and Blue = NW) as described in details in Ref [13], and shown in Fig. 3c. This also allows the quantification of the Pitsch, KS, NW and GT orientation relationships (Table 2). In each prior austenitic grain the four CPP packets are identified and colored in blue, yellow, red and green (Fig. 3d). The quality of the parent grain reconstruction is checked by plotting the pole figures with the calculated orientations of the austenitic grains and the experimental orientations of the martensitic grains they contain, and by checking that the special features described in section 6.1 are present (Fig. 3e). The orientations of the directions normal to the calculated {557} planes are plotted such that the reader can check that the traces of the habit planes are indeed perpendicular to these directions. The histogram of misorientations between the bcc grains is also established (Fig. 3f) and compared with literature. Before presenting the results on the habit planes, few words must be said about the block morphology. The intricate low-misoriented laths (KS variants) that constitute a block can be observed by TEM, but in SEM these laths are too small and internally deformed to be visible, and only the "average" block appears in the EBSD maps. However, by looking at the EBSD images, as shown in Fig. 3 and in the other EBSD maps, the blocks seem to be constituted of parallel regions. These regions are too long to be the laths or the sub-blocks. In addition, we noticed that some blocks appear as wide plates when the parent grains are cut on surfaces close to a $\{111\}_\gamma$ plane, as it is the case of the red bcc region in Fig. 3a inside the parent grain noted PG17 in Fig. 3b. More examples are shown in **Supplementary Material 2** (see the regions marked by the white arrows in Fig.S2.3). These regions can thus be interpreted by considering that a block is made of parallel foliated plates. The elongated regions in the EBSD maps are the traces of these foliations at the sample surface. In the rest of the paper, it is assumed that the traces of the habit planes of the blocks should be parallel to the traces of these plates.

The trace of the habit planes of the blocks in each prior fcc austenitic grains are automatically plotted in the EBSD maps and compared with the morphologies of the bcc blocks. The results for the 1018 steel presented in Fig. 3 are shown in Fig. 4a and b, for conditions (a) and (b), respectively. In most of bcc grains there are two possible habit planes; they are marked by fine white lines. If none of two candidates fit with the morphology of the block, i.e. is not parallel to the elongated direction of the block, the closest line is marked by bold lines. The grains where neither condition (a) or (b) is satisfied are marked by white disks. They correspond to regions where the reconstruction failed (white area in Fig. 3b) or where the calculated orientations of the parent grains is not correct (black area in Fig. 3c). By considering the Fig. 4a and b, it is concluded that none of the two cases (a) or (b) totally fits with the morphologies of the blocks. Condition (a) is checked for most of the blocks, but it fails for some large blocks, and condition (b) fails more often than condition (a), but for the small blocks. Nevertheless, we have checked that all the blocks have for habit plane a $\{557\}_\gamma$ plane. This was done by plotting in each prior austenitic grains the twelve $\{557\}_\gamma$ planes, and by manually translating those found to be parallel to the long directions of the blocks contained in this parent grain. The line length is then modified in order to be more or less proportional to the block length. Some lines are duplicated to be placed on different blocks. The traces that are not used are deleted. As shown in Fig. 4c, all the blocks could be identified to a $\{557\}_\gamma$ planes, even those which do not satisfy the conditions (a) and (b). The fit is always very good, within a tolerance better than 3°. In general, the number of different habit planes needed to fit with all the blocks in the large prior austenitic grains is between eight and ten. The same sequence of treatment, i.e. parent grain reconstruction, automatic trace analysis, manual trace positioning etc., was reproduced to treat another EBSD map acquired on the same sample. The result is given in Fig. 5. Here again, condition (a) was found to occur more often than condition (b), but the fit is not always good, and three areas of important discrepancies between the morphologies and the traces plotted according to the conditions (a) or (b) were marked by dashed rectangles: region A (PG6), region B (PG21) and region C (PG1). It was checked that the reconstruction was correct, and that, for regions A and B, the traces correspond to $\{557\}_\gamma$ planes, independently of condition (a) or (b), as shown in Fig. 6. The unique exception is the region C, where no agreement with $\{557\}_\gamma$ planes could be found (Fig. 7a). On the other hand, among the other types of habit planes usually reported in literature, we found a very good fit with the $\{225\}_\gamma$ planes (Fig. 7b).

Other EBSD maps have been acquired and treated on a Eurofer steel and a EM10 steel in their as-quenched states. The percentage of ORs (KS, NW, Pitsch and GT) and the histograms of disorientations are very close in these steels (Table 2, and **Supplementary Material 3**). The conditions (a) and (b) for the habit planes were also evaluated, and here again, the agreement is quite correct, but neither (a) nor (b) was found to fit perfectly the morphologies of the all blocks, even if a manual positioning shows that all the blocks match one of the $\{557\}_\gamma$ planes. An example of analysis on Eurofer is given in **Supplementary Material 4**. Eight maps were analyzed on the 1018, Eurofer and EM10 steels, representing more than fifty prior austenitic grains and more than three hundreds blocks. All blocks could be identified to a {557}$_\gamma$ plane, even when condition (a) or (b) fails. The only exception was the austenitic grain PG21 of Fig. 6 where the habit planes were found compatible with {225}$_\gamma$ planes.

## 7. Discussion

Let us do a short step backward that will be useful for the rest of the discussion. After having explored different alternative models for fcc-bcc martensitic transformation, such as the double-step model [22] and the one-step model based on Pitsch OR [23], we have orientated our researches for the last years toward a one-step model based on the KS OR [24]. The two-step model should apply to steels in which the intermediate hcp phase clearly, as it is the case in FeCrNi and FeMn steels [37][38], and the one-step Pitsch model is probably relevant for nitrogen iron alloys in which Pitsch OR is observed [39], but we came to seriously consider that for a vast majority of steels, the distortion that leads to KS OR is the "natural", "spontaneous" distortion and that the other orientations (Pitsch, NW, GT) result from this KS distortion due to the back-stresses induced in the surrounding matrix [24]. The KS OR was also found to be a crucial link between the other displacive transformations involving the fcc, bcc and hcp phases [28]. Thus, we continue considering that KS is the natural distortion in the low-carbon steels, and that the tendency toward the NW OR results from the coupling of the two-low misoriented variants in the blocks.

### 7.1. Confronting the experimental results on the {557} habit planes with the model

Experience shows that lath martensite are grouped by packets, and that packets are divided into three blocks, and each block is constituted by a pair of low-misoriented KS variants. As mentioned in introduction, the current view about the habit planes on the low carbon steels is that they are the {557}$_\gamma$ planes that contain the parallel dense directions habit plane. As the two variants of a block do not share their common close packed directions, it is assumed that one KS variant in the pair predominates the other and imposes its habit plane to the block. Contrarily to this current hypothesis, we have considered in our theoretical approach that the two low-misoriented KS variants play symmetric roles in their block, and we have shown that the average lattice distortion is exactly the one expected from the NW OR, see equation (18). Then, the calculations have shown that one of the three planes untilted by the NW distortion matrix is the plane $(1\ \sqrt{2}\ 1)_\gamma \approx (575)_\gamma$. This plane is at 9° far from the common close packed plane $(111)_\gamma$ // $(110)_\alpha$ but does not contain the common close-packed directions of any of the two KS variant of the block. The difference between our model and the current one is schematically illustrated in the case of one packet in Fig. 8. The three blocks are noted by their average NW orientations, i.e. $\alpha_1^{NW}$, $\alpha_2^{NW}$ and $\alpha_3^{NW}$, and the three habit planes are specified by the colors red, green, and blue, for the $(575)_\gamma$, $(755)_\gamma$ and $(575)_\gamma$ planes, respectively. Depending on the model chosen, the habit planes are attributed to different blocks. Clearly, the current view is dissymmetric and depends on the predominant variants chosen in the blocks. A 3D representation is given in Fig. 9a and b. The habit planes of the laths inside the blocks are added as darker bands in Fig. 9c with the current model, and in Fig. 9d with our model. It has been proven in the past TEM studies (see section 1) that the laths have a habit planes that contain the common dense direction; this plane could be {557}$_\gamma$ planes, as found by Morito et al. [11], but it could also be {225}$_\gamma$ planes, which are the planes untilted by the KS distortion [24]. We estimate that the experimental TEM studies are not precise enough to correctly distinguish these two cases. The habit planes used for the KS laths in Fig. 9d are $\{225\}_\gamma$ planes, but a very similar figure would be obtained with $\{557\}_\gamma$ planes. With the current view, the habit planes of one of the two KS laths is exactly the habit plane of the block, and the dissymmetry is created by the choice of this predominant variant. This dissymmetry is not present in our model.

In order to clarify the exact type of habit plane for each block, we have acquired and treated EBSD maps were on various low-carbon steels. The condition corresponding to the current view and to that of our model, i.e. the conditions (a) and (b) of section 6.2, respectively, were tested by plotting the traces of the expected habit planes and by comparing them with the martensite morphology. The condition (a) is more often respected than condition (b), but some very large blocks, such as the red one in the PG 17 of Fig. 4, were clearly shown to follow condition (b) and not (a). A manual positioning showed that all the blocks could be indexed very satisfactorily as $\{557\}_\gamma$ if one does not take into account any condition on the bcc plane. The only exception concerns the parent grain PG 21 in the EBSD map of Fig. 7, which shows martensite with habit planes could be $\{225\}_\gamma$. The absence of strict condition on the exact indices of the $\{557\}_\gamma$ habit plane can be explained if one considers that a block is predominant in the packet and imposes its habit plane to the two other blocks. Therefore, we substitute the current hypothesis of a predominant KS variant on the block, by the other hypothesis of a predominant block on the packet. To be clearer, let use the projection view of Fig. 9 along the direction $[\bar{1}11]_\gamma$ given in Fig. 10. It is clear that if one block imposes its habit plane to the other blocks, for example if all the blocks have the green $(755)_\gamma$ habit plane, neither condition (a) or (b) can be fulfilled. So, the fact that the habit plane of one block becomes predominant impedes an experimental distinction between the models. However, we think that our model is closer to the set of results reported in literature because it implies an equi-ratio of the laths, as observed by TEM. The TEM images also show that the laths are not parallel to the blocks and are shorter than the long direction of their block, which is also in good agreement with the fact that the habit plane of the block does not contain the long directions of the laths.

### 7.2. Variant pairing in low and high carbon steels

In high-carbon steels, the KS distortion has for untilted plane a $\{11\sqrt{6}\}_\gamma \approx (225)_\gamma$ plane [24], and we proved that a combination of two twin-related KS variants makes this $(225)_\gamma$ plane fully invariant [26]. In the case of low-carbon steels, we have shown that the average of the KS distortions associated with the two low-misoriented variants in a block is a NW distortion and that the plane untilted by this distortion is $(1\ \sqrt{2}\ 1)_\gamma \approx (575)_\gamma$. However, contrarily to the $(225)_\gamma$ planes, this plane is not rendered fully invariant by the variant pairing (the directions in this planes change). The $\{575\}_\gamma$ habit planes were confirmed by EBSD analyses (Fig. 3-Fig. 6); only one exception was encountered, and was compatible with $\{225\}_\gamma$ habit planes (Fig. 7). Why do the habit planes change from $\{225\}_\gamma$ to $\{575\}_\gamma$ when the carbon content is decreased? A partial answer can be provided. The transition temperature is higher in the low-carbon steels, which allows accommodation by plasticity, whereas the lower transformation temperature in high-carbon steels obstructs dislocation plasticity and favors the full accommodation of the habit plane by association of twin-paired variants (dislocations are however observed far from the midrib but in lower quantity than in low-alloy steels). One can then imagine that it is energetically more favorable forming at high temperatures low-misoriented blocky variants and accommodate the important remaining incompatibilities in the untilted $(575)_\gamma$ habit plane by plasticity. In comparison with formation of isolated individual KS laths, it is probable that the low-misoriented variant pairing allows reducing the amount of this plastic deformation. Indeed, the calculations of the average atomic displacement according to the distortion matrix $\mathbf{D}_1^{NW}$ is 0.43 Å instead of 0.46 Å for the distortion $\mathbf{D}_1^{KS}$ (see also **note 1**). The gain is not very high but may be a factor promoting the pairing. Two possibilities of plastic accommodation can be imagined, either in the surrounding fcc matrix, as suggested in fig. 9 of [24], or in the bcc crystal during its formation, as imagined in fig. 2 of ref.[40]. The fact that the distortion is mostly accommodated by low-misoriented variant pairing and plasticity in low-carbon steels is in agreement with the observation of the continuum of orientations between the Pitsch, KS and NW orientations observed in these alloys [13][22][23], whereas the fact that distortion is mostly accommodated by twin-variant pairing in high-carbon steels is in agreement with a more discrete distribution of orientations involving little plasticity [41]. These considerations are also coherent with the observations that the average orientation relationship is in-between the NW, KS and Pitsch in low-alloy steels, and very close to KS in high-carbon steels [42]. The exact nature of plastic accommodation of the $\{557\}_\gamma$ habit planes in the low-alloy steels remains to be established. Mesoscopically, two rotational gaps, one of 5.26° around $[\bar{1}10]_\gamma = [\bar{1}11]_\alpha$ and one of 5.26° around $[111]_\gamma // [110]_\alpha$, should be accommodated, which can be realized with two wedge disclinations [43][44], that are $\mathbf{w}_A$ = (5.25°, $[\bar{1}10]_\gamma$) and $\mathbf{w}_B$ = (5.25°, $[111]_\gamma$), respectively. More details can be found in the Supplementary Material 4 of ref. [23]. These disclinations are supposed to be at the origin of continuous rotations **A** and **B** observed in the pole figures [22]. In addition, since most of the accommodation is located at the austenite/martensite interface and since the $(557)_\gamma$ habit plane is close to the $(111)_\gamma$, the accommodation of blocky martensite should be mainly obtained by the disclination $\mathbf{w}_B$. It is probable that this disclination is a wall of screw dislocations as those observed in TEM by Sandvik and Wayman [8], but more theoretical and experimental work is required to confirm this point.

### 7.3. Self-accommodation in individual packets

We would like to end the discussion by considering other self-accommodation effects. The formation of blocks by pairing low-misoriented variants is a way to reduce the strains in the surrounding matrix, but other groupings can be imagined. The case of association of two blocks (four KS variants) has been investigated, but no special interesting result could be extracted; the details were reported in **Supplementary Material 5**. More interesting is the association of the three blocks sharing the same dense plane $\{111\}_\gamma // \{110\}_\alpha$ and forming a CPP packet. Let us show that this configuration is sufficient to accommodate the strains at the maximum, leaving only the hydrostatic part due to the volume change. In addition to the pair $\alpha_1^{KS}$-$\alpha_6^{KS}$ forming the block $\alpha_1^{NW}$ by averaging, we consider the pair $\alpha_2^{KS}$-$\alpha_3^{KS}$ forming the block $\alpha_2^{NW}$, and the pair $\alpha_5^{KS}$-$\alpha_8^{KS}$ forming the block $\alpha_3^{NW}$, as indicated in Fig. 8. It is reasonably assumed that averaging the six KS distortions is equivalent to averaging the three NW distortions.

For matrices written in polar decomposition, we assume that the average of the product is the product of averages

$$\langle \mathbf{D}_1^{NW}, \mathbf{D}_2^{NW}, \mathbf{D}_3^{NW} \rangle = \langle \mathbf{R}_1^{NW}, \mathbf{R}_2^{NW}, \mathbf{R}_3^{NW} \rangle \langle \mathbf{B}_1, \mathbf{B}_2, \mathbf{B}_3 \rangle$$

Averaging three rotations linked by a three-fold symmetry $g_{\frac{2\pi}{3}}$ is a special case of the example 2 treated by Moakher [36] in which the rotation axes are coplanar. In such a case, the rotation angle of the mean rotation is zero; thus, $\langle \mathbf{R}_1^{NW}, \mathbf{R}_2^{NW}, \mathbf{R}_3^{NW} \rangle = \mathbf{Identity}$. In addition, since the matrices $\mathbf{B}_x$, $\mathbf{B}_y$ and $\mathbf{B}_z$ are diagonal in the same basis, their average is simply

$$\langle \mathbf{B}_x, \mathbf{B}_y, \mathbf{B}_z \rangle = \left( \mathbf{B}_x \mathbf{B}_y \mathbf{B}_z \right)^{1/3} = \begin{pmatrix} \frac{2^{5/6}}{\sqrt{3}} & 0 & 0 \\ 0 & \frac{2^{5/6}}{\sqrt{3}} & 0 \\ 0 & 0 & \frac{2^{5/6}}{\sqrt{3}} \end{pmatrix} \quad (20)$$

Consequently, the average of the six KS variants in a packet, or equivalently of the average of their three mean NW distortion matrices, is an isotropic matrix equal to $\frac{2^{5/6}}{\sqrt{3}}$ times the identity matrix. It is the best distortion matrix that can be obtained by accommodation; its determinant $\frac{4\sqrt{\frac{2}{3}}}{3} \approx 1.088$ gives the volume change from fcc to bcc with hard-sphere condition. If the real change of atom size is taken into account, the multiplicative coefficient is lower, close to 1.016. Anyway, the change of size of the iron atoms doesn't change the calculations and conclusion: the martensitic transformation can be self-accommodated with six KS variants (or three NW variants) which are those that constitute an individual packet. The formation of any of the four packets is sufficient to obtain a self-accommodation; there is no need to form the 24 variants as it is sometimes believed.

## 8. Conclusion

The low-carbon steels are constituted of lath martensite with a highly intricate microstructure. The 24 variants are grouped into four crystallographic packets of six KS variants that share the same common close-packed plane. Each packet is constituted of three blocks of pairs of low-misoriented KS variants. The fact that orientation relationship with the prior parent austenite is not strict but follows a gradient from KS toward NW is not well explained in literature, so as the fact that the habit planes deviate from the $\{111\}_\gamma$ close-packed planes to be actually close to $\{557\}_\gamma$. The present work tries to answer these questions with simple crystallographic arguments, without adjusting any free parameters, without requiring hypothetical lattice invariant shear modes, without testing hundreds of thousand cases. First, it is shown that the two KS distortions associated with the two low-misoriented variants in a block induce by averaging a NW distortion. It should be noted that the averaging formula we used respects the volume change, contrarily to linear formulae applied in other theories. The predicted habit planes are the planes untilted by this distortion, i.e. they are eigenvectors of the distortion NW matrix expressed in the reciprocal space. The three predicted habit planes are $\{111\}_\gamma$, $\{557\}_\gamma$ and $\{110\}_\gamma$. The $\{111\}_\gamma$ are observed in FeNi meteorites and the $\{557\}_\gamma$ are the planes usually reported in literature. The $\{557\}_\gamma$ habit plane we predict are those at 9° from the common close-packed plane, but contrarily to the current model, they do not contain the close-packed direction of any of the two KS variants of the block. In order to clarify this discrepancy, three low-carbon steels have been studied by EBSD. The parent grains were automatically reconstructed and the traces of the expected habit planes plotted. The analyses confirm that the habit planes of the blocks are $\{557\}_\gamma$, and that the habit plane of one block often dominates on the other blocks in a packet. This predominance impedes us to experimentally discriminate the models. However, our model seems more in agreement with the TEM observations reported in literature because it implies an equi-ratio of KS variants in the blocks. The {557}$_\gamma$ habit planes are not rendered fully invariant by the variant pairing; so it is highly probable that the accommodation of the residual strains in a block is obtained by plasticity, and more precisely by disclinations. In addition, the calculations prove that the martensitic transformation can be completely self-accommodated in individual packets by the co-formation of the six KS variants, or equivalently three composite NW variants; the resulting mean distortion is the hydrostatic matrix due to the fcc-bcc volume change.

This work is in the continuity of our efforts [22][23][24][26] to develop an alternative, or at least a physical complement to the PTMC. It confirms the hypothesis that only one matrix, the angular distortion matrix associated with the KS OR, is sufficient to explain both the {225}$_\gamma$ habit planes in high-carbon steels [24][26], and now the {557}$_\gamma$ habit planes in low-alloy steels. Although the irrational character of these habit planes was sensed for a long time, it is only now that real irrational values are proposed. According to our model, the {225}$_\gamma$ and {557}$_\gamma$ habit planes are in fact {11$\sqrt{6}$}$_\gamma$ and {11$\sqrt{2}$}$_\gamma$ planes, respectively. Our research now continues to better understand the {259}$_\gamma$ habit planes found in Fe-Ni-C steels.

**Acknowledgements:**

We acknowledge the Swiss National Science Foundation for his support (grant n°200021_159955). PX Group is also thanked for its financial support to the laboratory. We are also very grateful to our colleague Dr Yann de Carlan at CEA-Saclay who provided us the EM10 steel.

**Note 1:**

The values of the mean displacement per atom calculated in section 7.3. of Ref.[24] were based on a classical unit cell formed by the atoms with positioned in [0,0,0], [1,0,0], [0,1,0], [0,0,1], [1,1,0], [1,0,1], [0,1,1], [1,1,1], [1/2,1/2,0], [1/2,0,1/2], [0,1/2,1/2], [1/2,1/2,1], [1/2,1,1/2], [1,1/2,1/2]. The corner atoms (eight atoms) count for 1/8, and the atoms in the centers of the faces (six atoms) for ½; the unit cell contains 4 atoms. The average displacement is calculated by summing the pondered displacements of the 14 atoms and dividing the result by 4. The mean displacement was calculated with the Bain, Pitsch and KS distortion matrices. Unfortunately, we realized during the preparation of present paper, that the result obtained in Ref.[24] depends on the choice of the reference frame. This is because the origin [0,0,0] of the basis is a corner of the unit cell whereas it should be its center. The results obtained with a cell centered in [0,0,0], i.e. by translating the vectors by -[1/2,1/2,1/2] now give mean displacements that do not depend on the choice of the reference frame. The mean displacements recalculated with the centered unit cell are 0.347 Å for Bain, 0.461 Å for KS and 0.429 Å for Pitsch. Therefore, we were wrong when we said that the displacements associated with the Pitsch and KS distortion were lower than for Bain distortion. The rest of the paper [24] is not affected by this error, and the conclusion that the distortion associated with the KS OR was a good candidate to be the natural distortion and could explain the {225} habit planes in the high carbon steels is not changed, as if was later fully confirmed in Ref. [26].

**Note 2:**

We take the opportunity of the paper to correct a typo error in the indices of the phases that appeared in equation (5) of Ref. [28] during the edition process. The orientation relationships should be read:

KS: $[110]_\gamma = [111]_\alpha$ and $(\bar{1}11)_\gamma // (\bar{1}10)_\alpha$

Burgers: $[111]_\alpha = [100]_\varepsilon$ and $(\bar{1}10)_\alpha // (001)_\varepsilon$

SN : $[110]_\gamma = [100]_\varepsilon$ and $(\bar{1}11)_\gamma // (001)_\varepsilon$

**Tables**

| Wt-% | Cr | Mo | W | C | N | V | Ni | Si | Mn | Ta |
|---|---|---|---|---|---|---|---|---|---|---|
| **1018** | - | - | - | 0.18 | - | - | - | - | 0.8 | - |
| **Eurofer** | 9.0 | - | 1.04 | 0.109 | 0.02 | 0.18 | 0.07 | 0.04 | 0.5 | 0.1 |
| **EM10** | 8.8 | 1.1 | - | 0.096 | 0.02 | 0.03 | 0.18 | 0.37 | 0.5 | - |

Table 1. Chemical composition of the three low-carbon steels used in the study.

| OR | Pitsch | KS | NW | GT |
|---|---|---|---|---|
| **1018** | 3 | 22 | 44 | 31 |
| **Eurofer** | 7 | 25 | 34 | 34 |
| **EM10** | 8 | 29 | 30 | 33 |

Table 2. Surface percentage of the different orientation relationships (ORs) determined from the analysis of the EBSD maps. An error bar of 15% is roughly estimated from the different measurements.

**Figure captions:**

Fig. 1. Schematic representation of the KS and NW variants. (a) 3D view with the $\{111\}_\gamma$ faces forming a regular pyramid, and the $\{110\}_\alpha$ // $\{111\}_\gamma$ planes of the 24 KS variants represented by rectangles. The KS OR imposes that for each variant a close-packed $<111>_\alpha$ direction is parallel to a $<110>_\gamma$ direction, i.e. a diagonal of the rectangle is parallel to an edge of pyramid. The packets are formed by the 6 variants lying on the same $\{111\}_\gamma$ face. The special twin-related pairs and the blocky pairs are colored in yellow and marked by the red and blue ellipses, respectively. (b) Planar view of the packet $(111)_\gamma$ // $(110)_\alpha$ plane showing one block constituted of the KS variants $\alpha_1^{KS}$ and $\alpha_6^{KS}$, with the average block orientation noted by the $\alpha_1^{NW}$ variant. (c) Representation of the angular distortion part of the KS distortion associated with the two KS variants $\alpha_1^{KS}$ and $\alpha_6^{KS}$. Instead of using rectangles, the variants are represented by triangles formed by the dense directions $<111>_\alpha$. The $(111)_\gamma \rightarrow (110)_\alpha$ fcc-bcc transformation occurs by opening the angle between these directions by 10.5°. In the KS OR, only one directions is rotated while the other remains parallel to a dense direction of austenite.

Fig. 2. Simulated pole figures of the directions $<110>_\alpha$ of the 24 KS variants with their continuous paths [22], with (a) the $<111>_\gamma$ parent directions, and (b) the $<557>_\gamma$ parent directions.

Fig. 3. EBSD map of a 1018 steel in as-quenched state. (a) Daughter bcc grains, (b) reconstructed austenitic grains, (c) orientation gradients in the bcc grains in RGB coloring (Red = Pitsch, Green = KS, Blue = NW) [13], (d) Close-packed plane packets in each prior austenitic grains represented with four colors, (e) pole figures on some parent grains with the $<110>_\alpha$ direction in blue, and the calculated $<557>_\gamma$ directions in red. (f) Histogram of disorientations between the daughter bcc grains in the whole map.

Fig. 4. Determination of the habit planes in the EBSD map of Fig. 3 by (a) automatically plotting for all the bcc grains the traces of the planes $\{156\}_\alpha$ // $\{557\}_\gamma$, (b) automatically plotting for all the bcc grains the traces of the planes $\{750\}_\alpha$ // $\{557\}_\gamma$, and (c) manually placing the traces of the planes $\{557\}_\gamma$ that fit with the morphology of the blocks. The white dots in (a) and (b) corresponds to bcc grains where neither condition (a) nor (b) works.

Fig. 5. EBSD map of a 1018 steel in as-quenched state. (a) Daughter bcc grains, (b) reconstructed austenitic grains, (c) orientation gradients in the bcc grains in RGB coloring (Red = Pitsch, Green = KS, Blue = NW), (d) CPP packets in each prior austenitic grains, represented with four colors.

Fig. 6. Regions A and B of the EBSD map of Fig. 5 (corresponding to PG6 and PG1, respectively). Traces of the $\{557\}_\gamma$ planes manually placed where the fit with the morphology of the blocks is correct.

Fig. 7. Region C of the EBSD map of Fig. 5 (PG21). (a) Traces of all the $\{557\}_\gamma$ planes, (b) Traces of the $\{225\}_\gamma$ planes manually placed where the fit with the morphology of the blocks is correct.

Fig. 8. Schematic view of a packet with the three blocks constituted by pairs of KS variants placed at the corners of the $(111)_\gamma$ triangle and their habit plane represented by a colored rectangle,

(a) according to the current view, (b) according to our model. The names of the KS variants are placed along their close-packed direction shared with the austenite. The average orientation of a block is noted its NW orientation. The blocks $\alpha_1^{NW}$, $\alpha_2^{NW}$, $\alpha_3^{NW}$ are constituted of the pairs $\alpha_1^{KS}$-$\alpha_6^{KS}$, $\alpha_2^{KS}$-$\alpha_3^{KS}$, $\alpha_5^{KS}$-$\alpha_8^{KS}$, respectively. The green, red, blue, colors are attributed to the $(755)_\gamma$, $(575)_\gamma$ and $(557)_\gamma$ habit planes, respectively, independently of the blocks and their KS variants. The configuration (a) is clearly dissymmetric.

Fig. 9. Representation of the $\{557\}_\gamma$ habit planes in the $(111)_\gamma$ packet and their associated blocks. (a) 3D view of the $(755)_\gamma$, $(575)_\gamma$ and $(557)_\gamma$ habit planes plotted in the unit cube, colored in green, red, blue, respectively. (b) Other axis of view. The intersections of the habit planes are three $\langle 5,5,12 \rangle_\gamma$ lines colored in red. The habit planes are attributed to the blocks: (c) according to the current model, or (d) according to our model. The KS laths are colored with the same color as their block, but in darker. In the current model (c), one of the two KS laths is exactly parallel to the habit plane of their block (predominant variant) and is not represented here to gain graphical visibility.

Fig. 10. Projection along the [011] direction of a $(111)_\gamma$ packet. (a) 3D representation of the projection. The habit planes are attributed to the blocks: (a) according to the current model, and (b) according to our model. The traces of the blocks and laths are represented by the (red, blue, green) colored rectangles and blue lines, respectively. The dissymmetry of the configuration (b) is apparent when considering the traces of the KS lath variants (the blue lines).

**Figures**

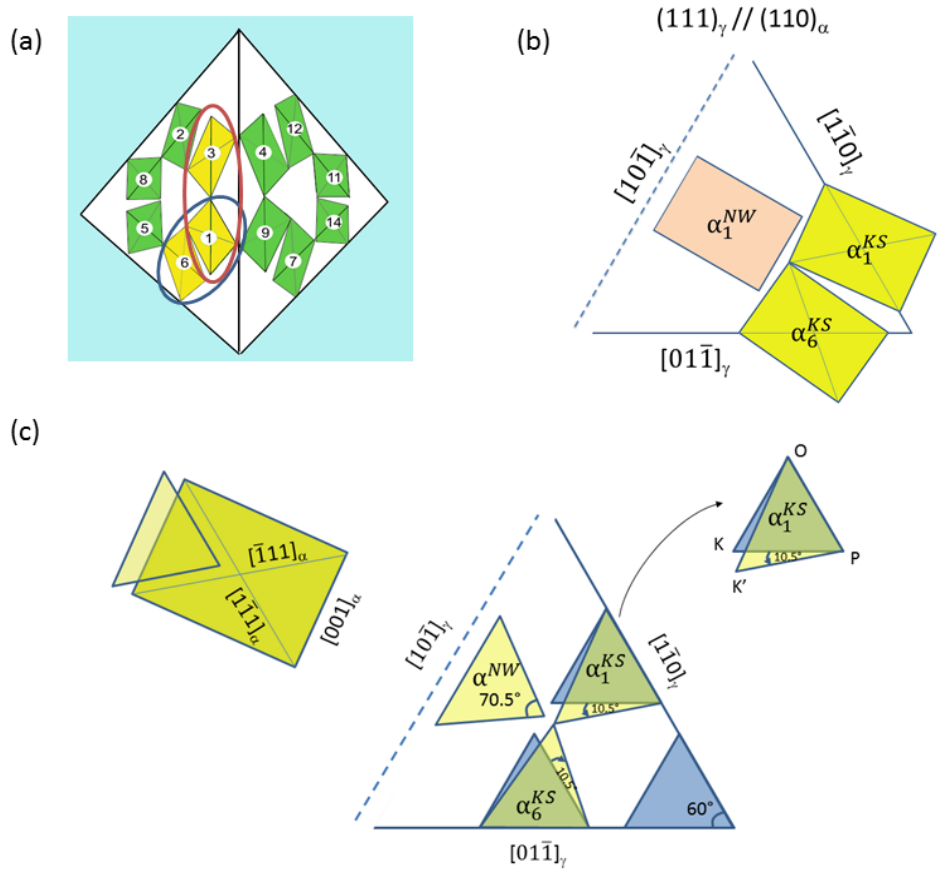

Fig.1.

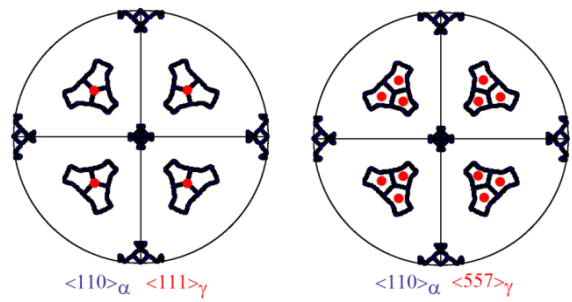

Fig. 2

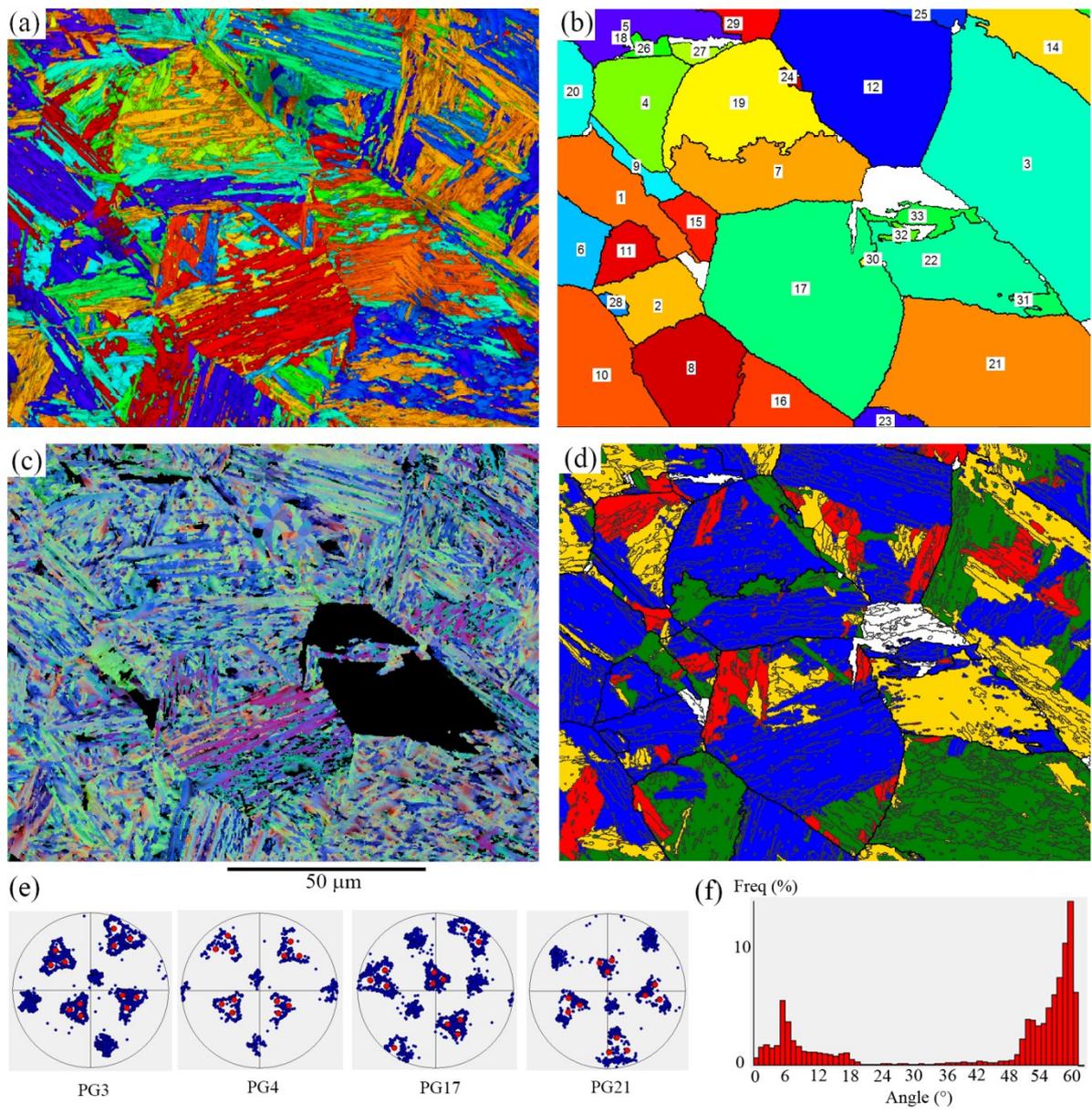

Fig. 3

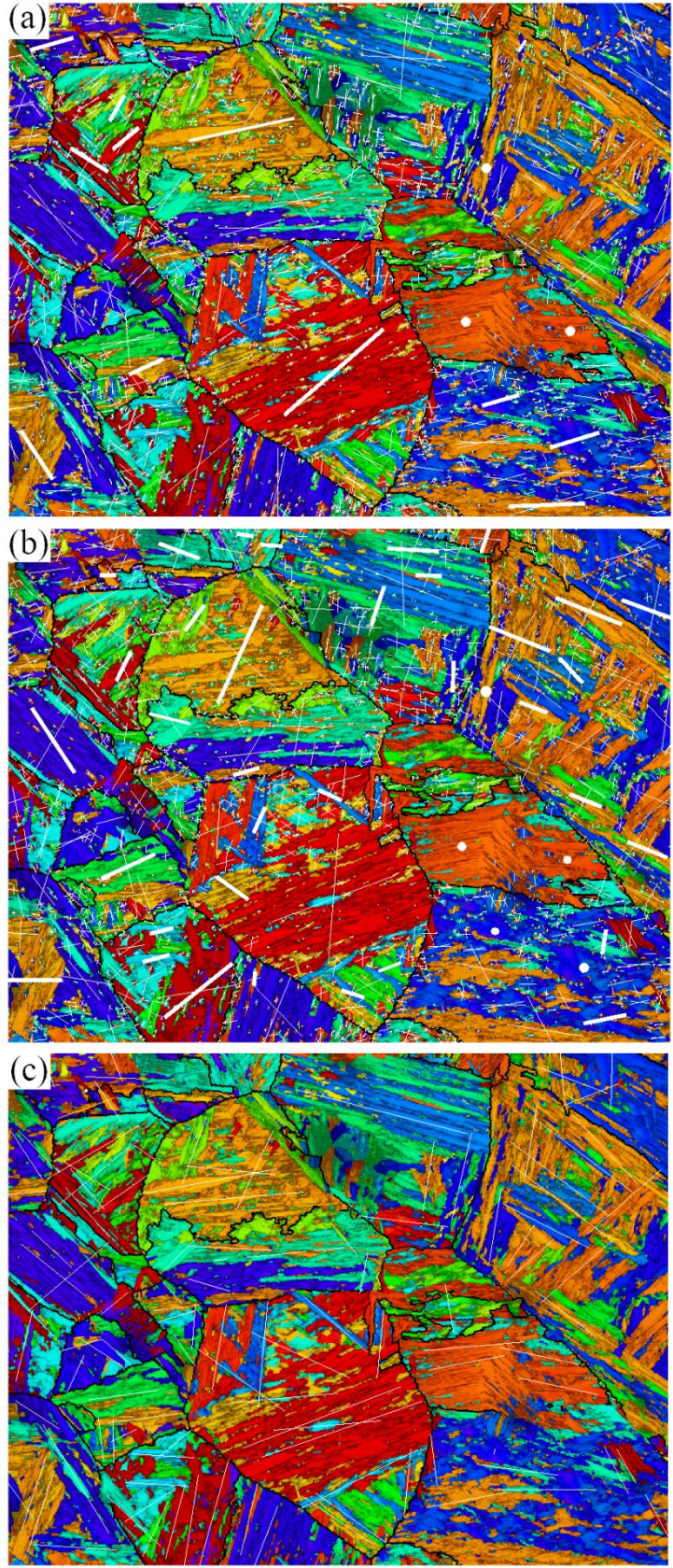

Fig.4

Fig. 5

Fig. 6

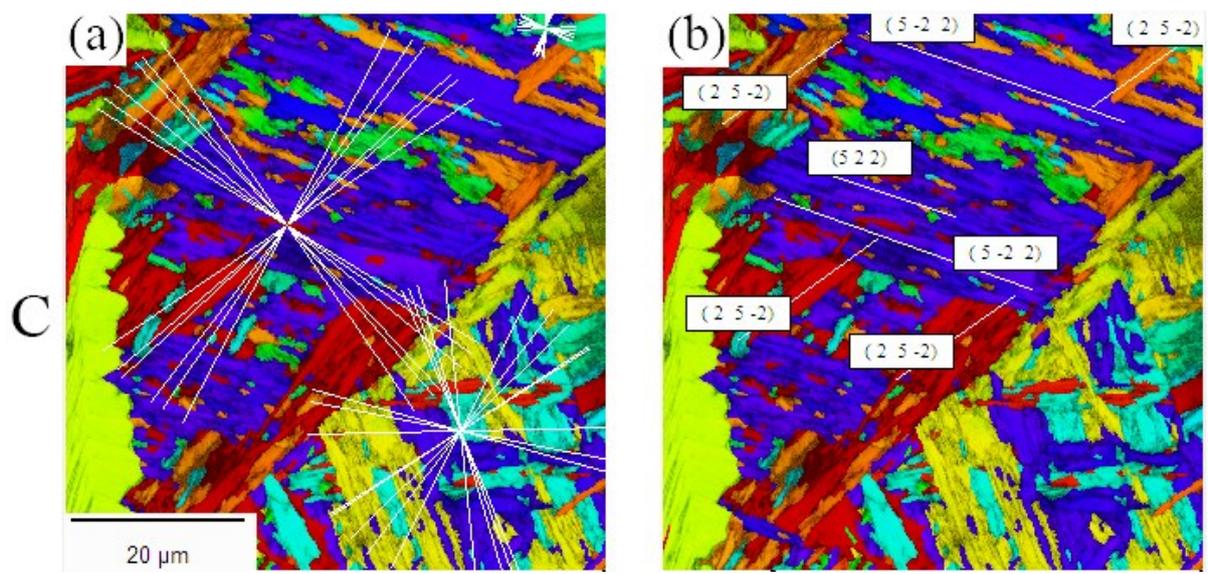

Fig. 7

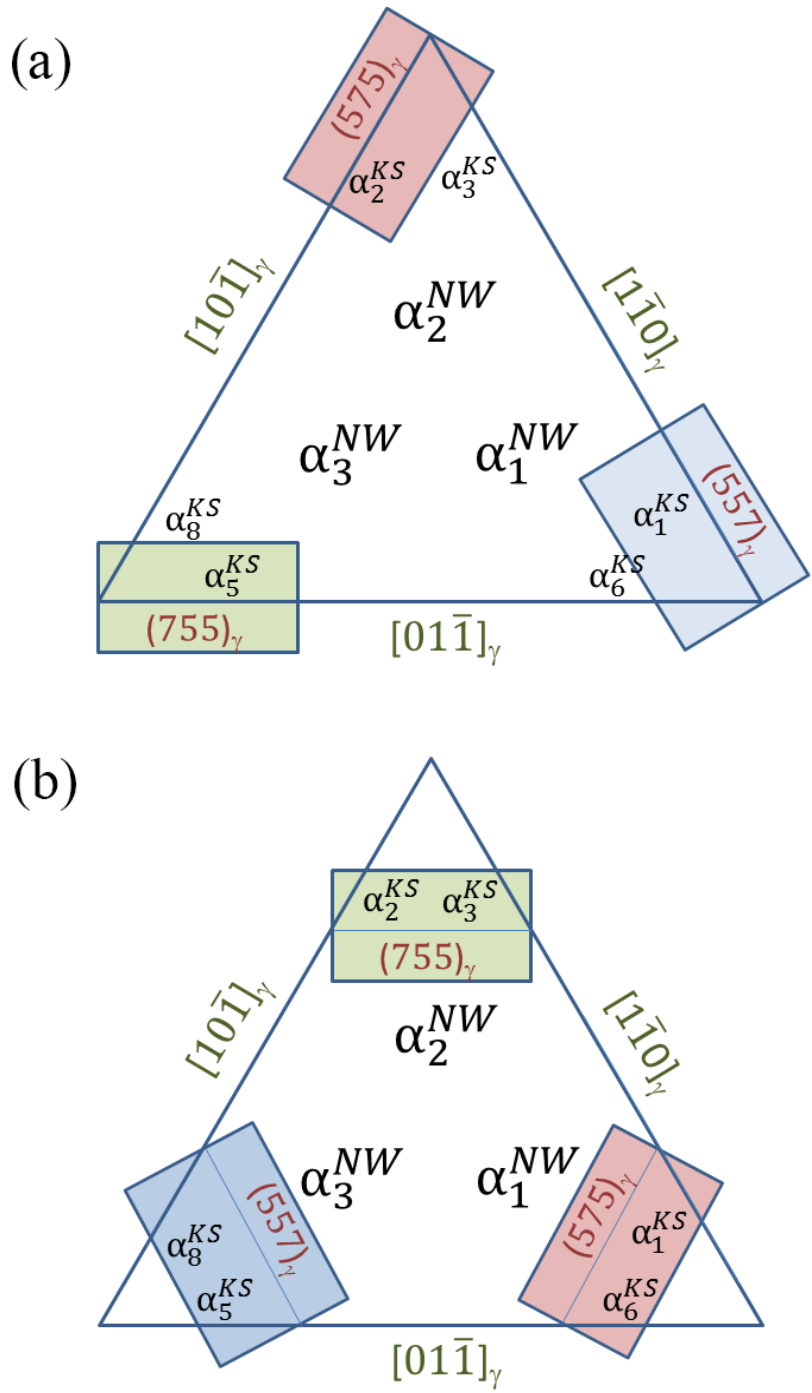

Fig.8

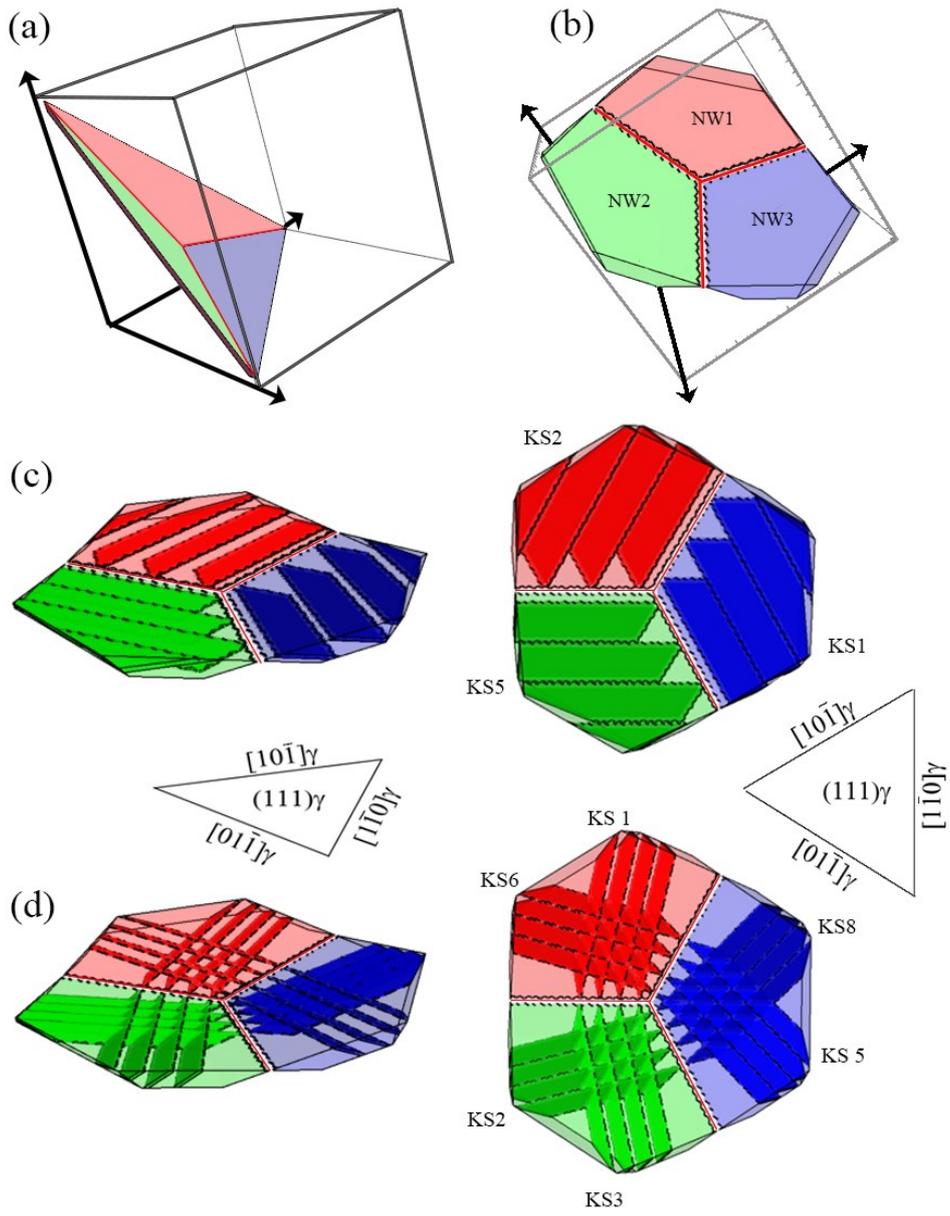

Fig.9

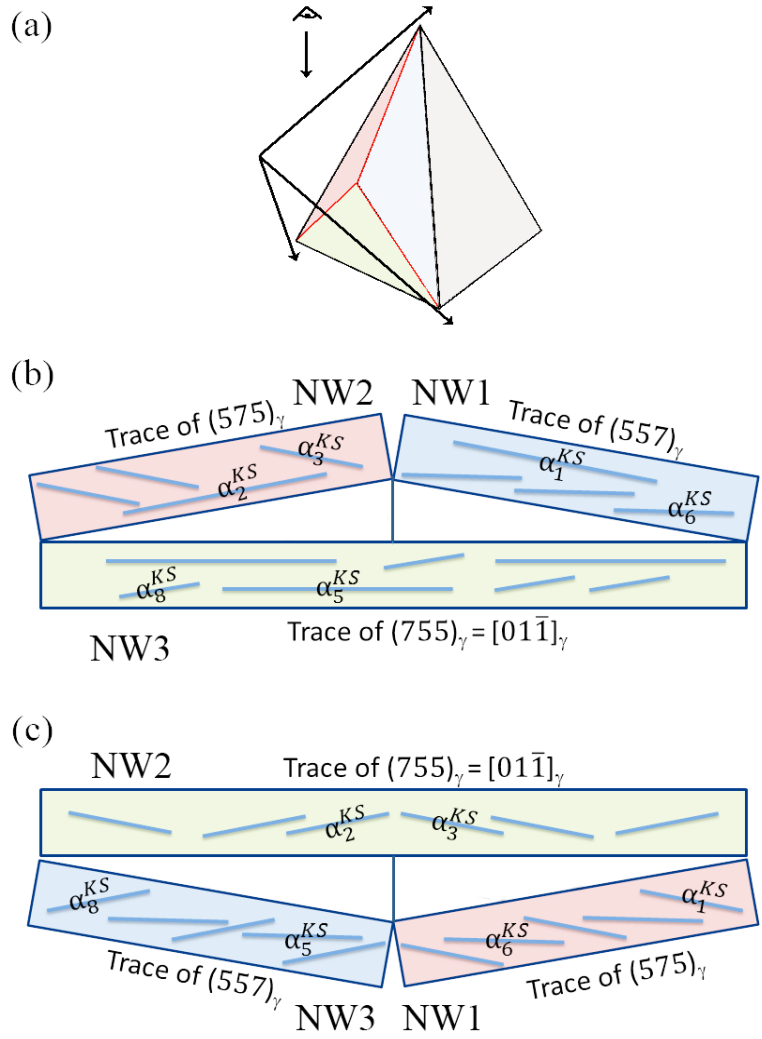

Fig.10

**Reviewer'report**

*with the corresponding author's responses written in green.*

Reviewer #2: This manuscript deals with a study of the morphology and crystallography of lath martensite in low carbon steels. The introduction begins with the authors' interpretation of the current understanding of the morphological aspects of lath martensite, moves on to a somewhat critical review of the phenomenological theory of martensite crystallography (PTMC), then to the recent development of an "alternative model" to the PTMC by the corresponding author (Cyril Cayron), and ends with a description of the aims of the paper. These aims are to show that "average distortion of the two KS variants in a block" can be regarded as a distortion matrix based on the NW orientation relationship, that the new alternative model predicts the block habit plane is {575}F, even though this plane "does not contain any of the common close-packed directions of the two variants in the block, which is in apparent contradiction with the current view". Finally, Electron BackScatter Diffraction (EBSD) maps are used to reconstruct the original austenite grain orientations and compare the habit plane traces with the new theoretical predictions of a {575}F block habit plane. Following the introduction, there is a short paragraph on experimental methods, a lengthy mathematical description of the new model and its predictions and finally the experimental analysis that attempts to show that the results are consistent with these new model predictions.
To fully assess the significance and originality of this manuscript and to help put some of the confused and occasionally incorrect statements into perspective it is essential to summarise the current understanding of the morphology and crystallography of lath martensite in steels. Over the past 60 or 70 years there has been very extensive published experimental and theoretical work on this topic, some, but not all of which appears in the list of references. The culmination of all this work is probably best summarised in a pair of excellent recent papers by Morito et al - given as references [10] and [11] in the current manuscript. It is now accepted that the individual martensite laths in low carbon steels have a long direction along the <110>F, which is within 0 to 5o of <111>B, and a {557}F habit plane, which is some 8 to 10o away from the {111}F plane that is parallel to {110}B in the orientation relationship (OR). This OR can range from the KS OR to the NW OR, including the intermediate GT OR. Within a single parent austenite grain there are many approximately equiaxed shaped 'packets' of martensite laths. Each of these packets contains laths that all share a common parallel pair of close packed planes - i.e. a quarter of the 24 KS or GT OR variants. Hence there will be 6 possible variants in each packet and there must be three other packet arrangements to accommodate the other three groups of 6 variants, giving an absolute minimum of 4 packets per parent austenite grain. A packet is usually made up of three "extended parallel blocks", as shown in Figures 1 and 2 of reference [11]. In the lower carbon steels, each block contains a pair of {557}F lath variants with <110>F //<111>B long directions that lie in the common parallel pair of close-packed planes {111}F//{110}B for the specific group of variants in the packet.

*Till here, I agree.*

Hence the common parallel plane separating these blocks - their habit plane - should be {111}F //{110}B.

*Why should the habit plane of the blocks be the same as that of the laths? Why should they be (111)F and not (557)F ?*

At higher carbon contents the blocks tend to contain only laths of a single variant. The two types of parallel blocks are illustrated in Figure 10 of [10]. *Figure 10 of [10] is schematic illustration, not an experimental proof.* The habit plane of these parallel blocks was not specifically determined in the papers by Morito et al but previous work has indicated that the laths "tend to group themselves in plate-like arrays which delineate {111} planes" - see [25]. *Ref[25] is Jaswon and Wheeler's paper and it does not concern the {557} martensite.* The claimed significance of the manuscript is that the new theoretical treatment is a real alternative to the existing

Phenomenological Theory of Martensite Crystallography (PTMC), generates new theoretical predictions and that the EBSD mapping technique can provide reliable experimental evidence to verify these predictions.

*At least, we showed that the experimental evidences do not contradict the model.*

Let us begin with the "alternative model to PTMC [22][23][24]". Their "alternative model" is identical to the untilted plane analysis published by Jaswon and Wheeler more than 60 years ago - see reference [25]. All the Cayron et al model does is reinvent the wheel. The authors need to show far more respect for previous work and openly admit that their "alternative model to PTMC" is nothing more than an outright copy of Jaswon and Wheeler's 1948 approach. Please give credit where credit is due.

*We admit to have discovered the work done by Jaswon and Wheeler (1948) once we finished our calculations on the fcc-bcc transformation [24]. We knew the theory of martensite mainly from Bhadeshia's book "Worked examples in the geometry of crystals" and there is no mention to Jaswon and Wheeler's work in this book. It is true that their paper is mentioned by Kelly in the section "Early theories of martensite crystallography" of chapter 1 of the book "Phase Transformations in Steels" 2012, Woodhead Publishing Limited, volume 2, edited by Edited by E. Pereloma and D.V. Edmonds, but we did not yet read this book at that time. We were not aware of Jaswon and Wheeler's work before finding the distortion matrix associated with the KS OR. There is no plagiarism and we invite the reader to look at Jaswon and Wheeler's paper and at our paper [24] to get a better idea whether Ref[24] is an "outright copy" or not. Ref[24] followed a logical way, initiated in 2010 with two-step model [22], and later in 2013 one step with Pitsch [23]). In addition, the way we came to the conclusions are different; and even if some parts of the conclusions are the same, the work [24] give the analytical expressions of the atomic displacements during the transformation, which is not done by Jaswon and Wheeler. We think we made credit to them by referring to Jaswon and Wheeler in our work [24] and [26]. Our work agrees with their work and we note that today PTMC has diverged too much from their analysis when four matrices instead of one were introduced few years by Bowles and Mackenzie, and by Wechsler, Lieberman, and Read. If Jaswon and Wheeler's paper is a "wheel", we are happy to have rediscovered it, and we invite other scientists rediscover it to find alternatives to PTMC in the hope to build an non-phenomenological (i.e more physical) theory of martensite.*

The associated discussion of the merits of the new alternative model mainly attempts to criticise the PTMC and its application to lath martensite. Unfortunately, all this achieves is to reveal the authors limited understanding of the field. The manuscript argues that the Ross/Crocker double lattice invariant shear (LIS) model used to explain the {557}F habit plane in lath martensite is a "trial-and-error method and the high degree of freedom given to the LIS significantly reduce the "predictive character" of these PTMC approaches." Yet the supposedly new model being used in the current paper has to assume a specific observed orientation relationship between the austenite and martensite, while the PTMC does not require this assumption. In fact the orientation relationship is one of its predictions.

*We assume the KS OR and hard-sphere atoms, and nothing else. In PTMC (which version? simple shear? double shear? which LIS? which lattice parameters? Measured at Ms or Mf?), there are more than thousands free parameters in the PTMC. One should not be impressed by the fact that PTMC "predicts" some ORs: PTMC starts from the Bain OR, which is at 10° from the experimentally observed OR, and then imposes the existence of an invariant line, which makes it closer to KS. If one considers the strong internal misorientations (up to 10°) inside the lath martensite, "finding" an OR close one of the experimentally observed OR should not be considered as a confirmation of the PTMC. In addition, PTMC did not predict these very particular internal misorientations that we consider as traces of the transformation mechanism [22,23,24].*

Another statement that casts doubt on the authors' appreciation of previous work is "In addition, to our knowledge, there is no clear experimental confirmation that the LIS proposed by the different authors fit with experimental observations such as those reported in TEM by Sandvik and Wayman [8]." This is totally and utterly wrong. The primary LIS used in the PTMC analysis of lath martensite - reference [17] in the current manuscript - is entirely consistent with the dominant Burgers vector of the tangled dislocations observed within the laths by Sandvik and Wayman in the paper designated [8] in the manuscript. Furthermore if the authors had gone to the trouble of reading the second Sandvik and Wayman paper in the same journal, they would have found that the PTMC analysis in [17] was also totally consistent with the interface dislocations observed by Sandvik and Wayman. - see Figure 6 and associated text in [17].

*The sentence "In addition, to our knowledge, there is no clear experimental confirmation that the LIS proposed by the different authors fit with experimental observations such as those reported in TEM by Sandvik and Wayman [8]." referred mainly to the works [19-21]. Concerning Kelly's work [17], we agree that the case is slightly different, and that we should have been clearer. However, we don't think that Kelly's claim "The double shear version of the phenomenological theory […] explains ALL the known crystallographic characteristics of lath martensite" is justified. We also disagree with the verb "predict" used when what is "predicted" was actually experimentally observed before the model establishment. Kelly's model uses many experimental observations as input parameters. For example, it is clearly stated that the two LIS S1 and S2 are built from the TEM observations made by Sandvik and Wayman; consequently, the model only predicts the distance between the dislocations but not their existence. In addition, using experimentally observed dislocations imposes the direction of the LIS, but there is still infinity of planes to fully define the LIS. To our knowledge there is no experimental work that tries to confirm that the LIS planes used in Kelly's model are the planes of the dislocation pile-ups, which supposed that the planes really exist and can be identified by TEM (the dislocations assemblies shown by Sandvik and Wayman seem to be complex and intricate).*

In Section 4 the authors move on to what they describe as "a new method of averaging distortion matrices". Anyone familiar with the relationship between the KS and NW ORs would appreciate that two KS variants separated by 10.53o could be combined artificially to give the NW OR that lies between them - see for example Figure 5 of U. Dahmen, Orientation Relationships in Precipitation Systems, Acta Met., 30 (1982) 63-73. So, although the Moahker [36] formula for the geometric mean of two rotation matrices is very elegant, it is unnecessary. Just calculate the corresponding NW OR matrix R1NW, if you really want to follow this approach. But don't believe that this is anything more than an artificial representation of the situation inside a block containing two groups of laths that each follow a different KS OR variant.

*We agree that the result was expected from a geometrical point of view, but we are happy to have been able to prove it algebraically because this effort helped us to better understand how average distortion matrices, and proved that the usual "linear" mean is in general incorrect.*

The predictions derived from the authors copy of the Jaswon/Wheeler analysis for the untilted planes with the NW OR are OK.

*That's a good point because it is an important result of the paper. May we remark that Jaswon and Wheeler did not "analyze" (557) lath martensite? The reviewer should have said Jaswon-Wheeler "criterion" of habit plane prediction. That's good if this criterion could be used again because to our knowledge the PTMC makes no reference to this criterion, and only O-lattice theory uses it without, to our knowledge, "giving credit" to Jaswon and Wheeler. For example, there is no mention to Jaswon and Wheeler in G.C. Weatherly and W.Z-Zhang paper "The Invariant line and precipitate morphology in fcc-bcc systems" Metall Trans A 1994, 1865.*

Where this predictive exercise goes completely astray is when it states "that (111)γ cannot be a habit plane for martensite because it is the plane in which the fcc-bcc volume change is "localized". What does this mean? Is this just because their artificial lattice parameters lead to no distortions normal to (111)F, so that the volume change must be accommodated within the plane? If they had used real lattice parameters this would not have happened and their excuse for excluding (111)F as a potential block habit plane would no longer exist.

*The artificial lattice parameters mentioned by the reviewer are those corresponding to an ideal hard-sphere packing. We agree that it is just an approximation, an idealization of the system in order to do calculations of the atomic trajectories. The distortion matrix that we calculated in [24] shows that, as for fcc-fcc twinning, there no change of the distance between the dense plane (111)F that are transformed into (110)B and that the volume change is only due to the angular distortion of the surface formed by the dense directions in the (111)F plane. We think that if the "real" lattice parameters are taken into account, an important part of the volume change continues to be localized in this plane. It is important because this way of seeing martensite transformation is really different from the usual shear paradigm (or its multiple shear versions). This is an example of calculation and discussion given in [24] and not by Jaswon and Wheeler.*

What is more important, if they had carefully read Jaswon and Wheeler and done some calculations for the KS and GT orientation relationships, they would have discovered that in all these examples the {111}F plane that is parallel to {110}B is always an untilted plane.

*Well, this point clear to us; the fcc-bcc continuous model proposed in [24] is based on this assumption. We don't understand this comment.*

For the pair of laths making up a block, BOTH of the lath variants will have the same untilted {111}F plane, and their <110>F long direction lies in this plane. The obvious conclusion is that the untilted {111}F plane is the favoured candidate to be the habit plane of the blocks. In their determination to exclude the untilted {111}F plane as the block habit plane and pursue their obsession with demonstrating that the habit plane of the blocks is the unrotated plane (1 √2 1)F or (5 7 5)F, the authors effectively cast doubts on the careful previous work of Marder and Krauss [6]. They claim that "Marder and Krauss [6] measured these habit planes by optical microscopy, i.e. at magnifications hundred times lower than TEM; which means that the {557}γ habit planes were those of the blocks (and not those of the laths)."

*Does it mean that the reviewer thinks that the habit planes of the blocks are (111)F planes ?!!*

*Krauss in Ref [14] used a TEM images to say that "most of the laths have widths smaller than 0.5 μm, the resolution limit of the light microscope, and therefore cannot possibly be revealed by light metallography." By considering that the determination they made of the {557} habit planes was done by optical metallography, it can be concluded that the {557} habit planes can't be those of the laths. They are more probably those of the blocks or of the packets. The EBSD investigation reported in the manuscript confirms that the {557} habit planes are those of the blocks.*

Micrographs such as Figure 3(b) of [6], the electron micrographs in Figure 10 and the associated discussion shows very clearly indeed that Marder and Krauss were perfectly justified in concluding that the {557}F habit plane applied to the individual laths and not to the blocks.

*Let us show these figures:*

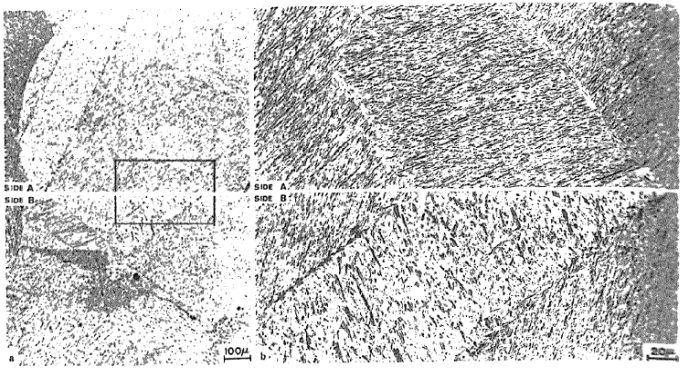

*Fig 3b, optical image*

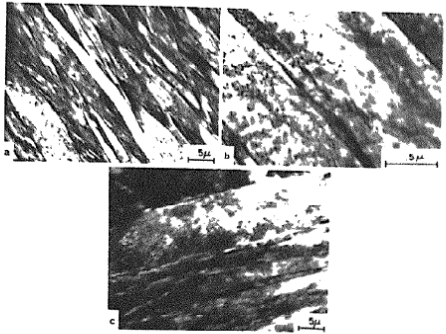

*Fig 10, TEM image*

*The determination of the habit plates was done on the large "laths" (today called "blocks") and not on the fine ones observed by TEM. When Marder and Krauss wrote their paper, the vocabulary was not established as it is today. The terms "blocks" and "packets" were used as synonymous for what is called today "packet", and the term "lath" was used as a generic term that includes both the fine laths (today called "laths") and the large laths (today called "blocks"). The assumption that lath and block/packets have the same habit plane was actually done later by Apple and Krauss in 1974. They presented TEM images (as the one shown below) and stated in conclusion that "the habit plane orientations of different laths in a packet are consistent with {557}A planes coupled to a single {111}A of the parent austenite".*

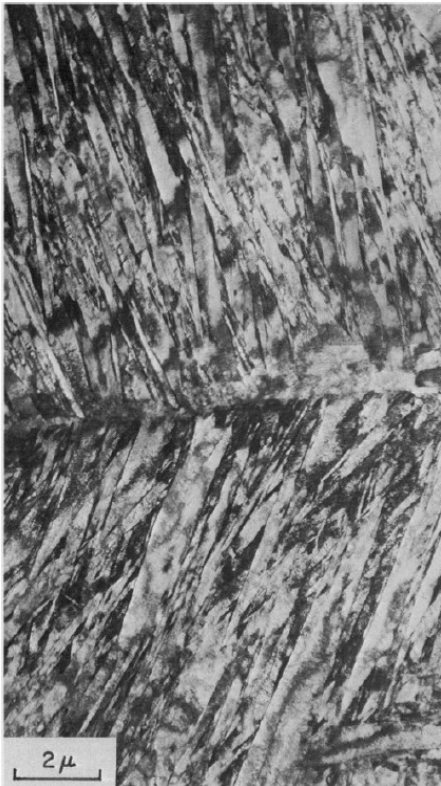

Fig. 2—Composite micrograph showing partions of two adjacent packets. Transmission electron micrograph.

*Unfortunately, the paper does not contain specific work on the determination of habit plane (by considering the SAED and by trace analysis of the TEM images) that supports this assumption. The reader can compare the scale bars of the TEM images and optical images to be convinced that fine laths ("lath") and large laths ("blocks") are different entities. Morito's work helped to classify these entities and avoid confusion in the terms.*

To claim otherwise is an insult to Marder and Krauss that is both unwarranted and unprofessional. The authors of the current manuscript go on to claim that in a later paper Krauss [14] "made the hypothesis that a group of laths with the same $\{557\}\gamma$ habit plane dominates on the other groups and imposes its habit plane" and they attempt to use this as support for their unjustified prediction that the block habit plane is (575)F. The manuscript authors should look carefully at Figure 10 of [11] where they will see that there are two types of blocks and the version referred to by Krauss [14] was the high carbon variety. This absolute obsession with advancing their prediction of (575)F as the block habit plane is totally and utterly unjustified, makes the rest of the manuscript essentially irrelevant and merely demonstrates the gaps in the authors understanding of the morphology and crystallography of lath martensite.

*There is no lack of respect to Marder and Krauss's work. We really admired the way they could determine the habit planes by optical microscopy, and there is no irony when we used the term "elegantly". However, we maintain our option that if the (fine) laths have really {557} for habit planes (we think that a deep TEM and trace analysis study would be required in order to confirm Morito's TEM work), nothing proves that they are the same as those of the blocks observed by Marder and Krauss by optical microscopy. Actually, assuming they are the same raises an important logical question: as the two laths in a block have their own (supposedly) {557} habit plane, which of two habit planes is the habit plane of the block? It seems that Krauss solved this problem by assuming that "Although there may be several crystallographic variants of laths in a packet of lath martensite, one variant or group of closely aligned variants may be dominant." [14]. Our model is based on*

*another assumption (equivalence of the two variants in the block, the distortion associated with a block is the average distortion the habit plane of the block is the plane untilted by the NW distortion).*

The portion of the manuscript devoted to the EBSD mapping and analysis is somewhat better. Here is an experimental technique that might have the degree of precision needed to distinguish between a block habit plane of (111)F as opposed to (575)F. This would severely test the accuracy of the method, because these two planes are only 8 to 10o apart and their traces in a surface could be even closer.

*Assuming (111)F habit planes for the blocks would imply that only four habit planes should be observed, which is clearly not the case. This argument was used by Marder and Krauss to prove that the habit plane of what should be nowadays called "blocks" are not (111)F.*

Adopting the two-surface habit plane determination approach used by Marder and Krauss would have helped to increase the accuracy of the habit plane determinations. Unfortunately, there is no attempt to distinguish between the two possible block habit planes, which is a great pity.

*We don't understand the remark. We treated the EBSD maps with two models in order to distinguish the two possible habit planes of the blocks. The attempt was done, even if we concede that it was not very successful and that the two models can't be distinguished experimentally for the moment. A multiscale analysis (TEM, ASTAR, TKD, EBSD, optics), probably coupled to FIB sample preparation, should be required for such a study. We hope to find time in the future this difficult experimental work.*

The two sets of Supplementary Material just generate more confusion. The meteorite data favours a {111}F block habit plane and the 1018 data merely demonstrates that the support for the {575}F block habit plane is not particularly convincing and does not effectively rule out a {111}F block habit plane.

*These supplementary materials show that the EBSD treatments allows the distinctions between (111)F habit planes, as it is the case in meteorites, and (557)F habit planes, as it is the case in low-carbon steels.*

**<u>Note:</u>** This manuscript can be cited by using its Arxiv reference number.